\documentclass[11pt]{article}
\usepackage{epsf}
\usepackage[nosort]{cite}
\newcommand{\nzmass}{91.19}
\newcommand{\nalphas}{0.118}
\newcommand{\nmcmc}{1.2}
\newcommand{\nmbmb}{4.2}
\newcommand{\nmtmt}{165}
\newcommand{\logmsms}{l_{ms}}

\newcommand{\dd}{{\rm d}}
\newcommand{\ddoverdd}[1]{{\dd\over \dd #1}}

\newcommand{\order}[1]{{\cal O}\left(#1\right)}

\newcommand{\nnb}{\nonumber}

\newlength{\captionwidth}
\newcommand{\msbar}{\overline{\rm MS}}
\title{
  \vspace{-2em}
  \begin{flushright}
    \bf\normalsize TTP00-09\\HET-BNL-00/8\\May 2000\\ hep-ph/0005139v3
  \end{flushright}
  \vspace{2em} Quartic mass corrections to $R_{\rm had}$ at
  $\order{\alpha_s^3}$}
\author{K.G. Chetyrkin$^a$, R.V. Harlander$^b$, and J.H. K\"uhn$^a$\\[2em]
  \it $^a$Institut f\"ur Theoretische Teilchenphysik,
  \it Universit\"at Karlsruhe,\\
  \it D--76128 Karlsruhe, Germany\\[.5em]
  \it $^b$Brookhaven National Laboratory, Upton, New York 11973} \date{}
\begin{document}
\maketitle 
\begin{abstract}
The total cross section for the production of massive quarks in
electron positron annihilation can be predicted in perturbative
QCD. After expansion in $m^2/s$ the quartic terms, i.e.\ those
proportional to $m^4/s^2$, are calculated up to order $\alpha_s^3$ for
vector and axial current induced rates. Predictions relevant for
charm, bottom and top quarks production are presented. The
$\alpha_s^3$ corrections are shown to be comparable to terms of order
$\alpha_s$ and $\alpha_s^2$. As a consequence, the predictions exhibit
a sizeable dependence on the renormalization scale. The stability of
the prediction is improved and, at the same time, the relative size of
the large order terms decreases by replacing the running mass
$\bar{m}(\mu)$ with the scheme independent invariant one $\hat{m}$. By
combining these results with the prediction for massless case and the
quadratic mass terms the cross section for massive quark production at
electron positron colliders is put under control in order $\alpha_s^3$
from the high energy region down to fairly low energies.
\end{abstract}

The total cross section for hadron production in
electron-positron annihilation is one of the most fundamental
observables in particle physics.  For energies sufficiently far above
threshold it can be predicted by perturbative QCD, and it is well
accessible experimentally from threshold up to the highest energies of
LEP and a future linear collider.  It allows for a precise determination
of the strong coupling $\alpha_s$ and, once precision measurements at
different energies are available, for a test of its evolution dictated
by the renormalization group equation. In many cases the cms energy is
far larger than the quark masses which motivated the original
calculations to be performed in the idealized case with the masses set
to zero from the start. In this limit, the results of
$\order{\alpha_s^2}$~\cite{CheKatTka79DinSap79CelGon80} and
$\order{\alpha_s^3}$~\cite{GorKatLar91SurSam91Che97} were obtained more
than two and nearly one decade ago, respectively (for a review
see~\cite{CheKueKwiPR}).

However, in a number of interesting cases quark masses do play an
important role~\cite{CheKueKwiPR}. For $Z$ decays into bottom quarks the
mass effects have to be included as a consequence of the extremely
precise measurements.  Bottom quark production at lower energies is
affected from production threshold up to a few tens of GeV. Other cases
of interest \cite{CheKue95,CheKueTeu35} are charm production between
roughly 5 and 10 GeV and, last but not least, top quark production at a
future linear collider~\cite{HarSte:eett}.

In two-loop approximation the full mass dependence of the cross section
has been evaluated since long and, exploiting the optical theorem, both
real and imaginary parts are available~\cite{KalSab55}. In three loop
approximation ($\order{\alpha_s^2}$) the corresponding results were
obtained only recently. Two methods have been used for this purpose: the
first one is based on the evaluation of a large number of terms for the
Taylor series of the polarization function $\Pi(q^2)$ at $q^2=0$ and an
appropriately chosen analytic continuation~\cite{CheKueSte96}. The
second one is based on the application of the large momentum expansion
(see~\cite{smirnov} and references therein), which provides an expansion
of $\Pi(q^2)$ in powers $(m^2/q^2)^n$, modulo
logarithms~\cite{CheHarKueSte97}. From the comparison between full
result and expansion one learns that the first few terms of the high
energy expansion provide a remarkably good description of the full
result, from high energies down to values of $2m/\sqrt{s} = 0.65$ --
0.75. The existence of the four quark threshold at $\sqrt{s} = 4m$ and
of a corresponding branching point for the polarization function
suggests that the high energy expansion diverges for $\sqrt{s}$ below
$4m$. Nevertheless, numerical studies \cite{CheHarKueSte97} as well as
qualitative arguments demonstrate that the sum of the first two or three
terms of the expansion can be trusted even down to $3m$ or even, with
some optimism, $2.5 m$ (where $m$ stands for the pole mass).

These considerations pave the way to a prediction of $R(s)$ including
the quark mass dependence to $\order{\alpha_s^3}$ along the following
route: In addition to the massless result the $m^2/s$ terms of
$\order{\alpha_s^3}$ have been calculated for the absorptive part nearly
a decade ago~\cite{CheKue90}. They were obtained by reconstructing the
logarithmic $\alpha_s^3m^2/s$ terms of $\Pi(q^2)$ from the full three
loop $\order{\alpha_s^2m^2/s}$ result of \cite{GorKatLar86} with the
help of the renormalization group equations. These are sufficient to
calculate the $m^2/s$ terms of the imaginary part in the time-like
region. The result is cast into a particularly compact form once
expressed in terms of the running mass $m(\mu^2)$, with the 't~Hooft
scale $\mu$ set to $\sqrt{s}$ throughout \cite{tHo73}, since this choice
eliminates all terms $\propto \ln(s/m^2)$. A generalization of this
approach has been formulated for the quartic mass terms in
\cite{CheSpi87,CheKue94} and was originally adopted for the calculation
of $\alpha_s^2 m^4/s^2$ terms. It is based on the operator product
expansion and the usage of the renormalization group equation to again
construct the logarithmic terms of the polarization function. In
addition to the anomalous mass dimension and the $\beta$--function the
anomalous dimensions of the operators of dimension four are required in
appropriate order.

This method also allows to determine the $\alpha_s^3m^4/s^2$ terms.  The
calculation can be reduced to the evaluation of massless propagators and
massive tadpole integrals, both at most in three loop approximation.
Details of the calculation will be given elsewhere \cite{CheHarKueprep}.
Compared to the case of the $m^2/s$ terms an important difference
arises: Even adopting the $\msbar$ definition (at scale $\mu^2=s$) for
the quark mass, logarithmic terms $\propto \ln(s/m^2)$ remain
in order $\alpha_s^2$ and above. The power of these logarithms,
however, is reduced by one. In fact, these logs may be also summed up
(see Refs.~\cite{BroGen84,CheSpi87}).

To predict the $R$ ratio, it is well justified to consider only one
quark as massive and to neglect the masses of the lighter quarks. The
heavier quarks decouple (apart from the tiny axial-vector singlet
contribution --- see below) and can therefore be omitted in the
calculations. The massive quark will generically be denoted by $Q$ in
the following, while $q$ refers to all the lighter quarks.

In the energy region where quark mass effects are relevant, charm and
bottom production is solely induced by the electromagnetic vector
current. Top quark production, however, which sets in above 350~GeV,
receives additional contributions from the axial current. Both vector
and axial-vector will be considered separately in the following.  The
prediction for both of them is conveniently split up as follows:
\begin{eqnarray}
R^{(v)} &=& 3\left( \sum_q v_q^2\,r_{q}
 +  v_Q^2\,r_{Q}^{(v)}
 +  r^{(v)}_{\rm sing}
 \right)\,,
\label{eq::Rdef}
\end{eqnarray}
and similarly for the axial-vector part ($v\to a$).  The sum runs over
all massless quark flavors. $v_{q/Q}$ ($a_{q/Q}$) is the coupling
constant to the light/heavy quarks in the vector (axial-vector) case.
For low energies only the electromagnetic current contributes, $v\to e$
and $a\to 0$. For high energies, both the electromagnetic and neutral
current pieces are relevant and have to be included with appropriate
weights (see, e.g., \cite{KueZerBL}).
$r_q$ and $r_Q^{(v/a)}$ represent the non-singlet contributions
arising from diagrams where the external currents are linked by a common
quark line.  They originate from two different types of diagrams: For
$r_q$ the external current couples to massless quarks; the massive
quark then only appears through its coupling to virtual gluons.
$r_q$ is the same for external vector and axial-vector currents.
On the other hand, $r_Q^{(v/a)}$ corresponds to diagrams where the
external current couples to the massive quark.  $r_{\rm sing}^{(v/a)}$,
finally, comprises massless and massive singlet contributions, where
either of the external currents is coupled to a separate closed quark
line.

Both $r_q$ and $r_{Q}^{(v/a)}$ are written as series in $m_Q^2/s$,
where $s$ is the cms energy and $m_Q$ is the $\msbar$ mass of the
quark $Q$:
\begin{equation}
  r_{q} = r_0 + r_{q,2} + r_{q,4} 
  + \ldots\,,\qquad
  r^{(v/a)}_{Q} = r_0
  + r^{(v/a)}_{Q,2} + r^{(v/a)}_{Q,4} + \ldots\,.
\end{equation}
$r_0$ denotes the massless approximation, while $r_{q,n}$ and
$r_{Q,n}^{(v/a)}$ are the mass terms of order $m_Q^n$. If not stated
otherwise, the renormalization scale $\mu^2=s$ is adopted below.

Denoting $n_f$ the number of active flavors, the massless terms are
given by (all the following formul\ae{} are valid up to
$\order{\alpha_s^3}$ unless indicated otherwise)
%
%
\begin{eqnarray}
  r_{0} &=& 1 + {\alpha_s\over \pi} + \left({\alpha_s\over
    \pi}\right)^2\,\bigg[ {365\over 24} - 11\,\zeta_3 + n_f\,\bigg(
  -{11\over 12} + {2\over 3}\,\zeta_3 \bigg) \bigg]
                                                              \label{eq::r0}
\\&&\mbox{\hspace{0em}}
      + \left({\alpha_s\over \pi}\right)^3\,\bigg[
          {87029\over 288} 
          - {121\over 8}\,\zeta_2 
          - {1103\over 4}\,\zeta_3 
          + {275\over 6}\,\zeta_5
\nnb\\&&\mbox{\hspace{1em}}
          + n_f\,\bigg(
              -{7847\over 216} 
              + {11\over 6}\,\zeta_2 
              + {262\over 9}\,\zeta_3 
              - {25\over 9}\,\zeta_5
              \bigg) 
          + n_f^2\,\bigg(
              {151\over 162} 
              - {1\over 18}\,\zeta_2 
              - {19\over 27}\,\zeta_3
              \bigg) 
          \bigg]
\nnb\\
 &\approx&
      1
      + {\alpha_s\over \pi} 
      + \left({\alpha_s\over \pi}\right)^2\,\Big(
         1.98571 - 0.115295\,n_f \Big)
         \nnb\\&&\mbox{}
      + \left({\alpha_s\over \pi}\right)^3\,\Big(
         -6.63694 - 1.20013\,n_f - 0.00517836\,n_f^2\Big)
\nnb\,.
    \end{eqnarray}
%

The quadratic mass corrections, separated according to
(\ref{eq::Rdef}), read~\cite{CheKue90,CheKue97}:
%
\begin{eqnarray}
 r_{q,2} &=&
{m_Q^2\over s}\,\left({\alpha_s\over \pi}\right)^3\,\bigg[
          -80 
          + 60\,\zeta_3
          + n_f\,\bigg(
              {32\over 9} 
              - {8\over 3}\,\zeta_3
              \bigg) 
          \bigg]\\
 &\approx&
       {m_Q^2\over s}\,\left({\alpha_s\over \pi}\right)^3\,
\bigg[-7.87659 + 0.35007\, n_f\bigg]\,,
\nonumber\\[1em]
 r_{Q,2}^{(v)} &=&
      {m_Q^2\over s}\,\bigg[
          12\,{\alpha_s\over \pi} 
          + \left({\alpha_s\over \pi}\right)^2\,\bigg(
              {253\over 2} 
              - {13\over 3}\,n_f
              \bigg) 
\\&&\mbox{\hspace{0em}}
          + \left({\alpha_s\over \pi}\right)^3\,\bigg(
              2442 
              - {855\over 2}\,\zeta_2 
              + {490\over 3}\,\zeta_3 
              - {5225\over 6}\,\zeta_5 
\nonumber\\&&\mbox{\hspace{1em}}
              + n_f\,\Big(
                  -{4846\over 27} 
                  + 34\,\zeta_2 
                  - {466\over 27}\,\zeta_3 
                  + {1045\over 27}\,\zeta_5
                  \Big)
              + n_f^2\,\Big(
                  {125\over 54} 
                  - {2\over 3}\,\zeta_2
                  \Big) 
              \bigg)
              \bigg]\nnb\\
          &\approx&
          {m_Q^2\over s}\,\bigg[ 12\,{\alpha_s\over \pi} + 
   \left({\alpha_s\over \pi}\right)^2\,\Big(
      126.5 - 4.33333\,n_f\Big)
\nnb\\&&\mbox{}
 + \left({\alpha_s\over \pi}\right)^3\,\Big(
1032.14 - 104.167\,n_f + 1.21819\,n_f^2\Big)
\bigg]\,,
\nnb\\[1em]
 r_{Q,2}^{(a)} &=&
      {m_Q^2\over s}\,\bigg[
          -6 
          - 22\,{\alpha_s\over \pi} 
          + \left({\alpha_s\over \pi}\right)^2\,\bigg(
              -{8221\over 24} 
              + 57\,\zeta_2 
              + 117\,\zeta_3
\nonumber\\&&\mbox{\hspace{1em}}
              + n_f\,\Big(
                  {151\over 12} 
                  - 2\,\zeta_2 
                  - 4\,\zeta_3
                  \Big) 
              \bigg) 
\nonumber\\&&\mbox{\hspace{0em}}
          + \left({\alpha_s\over \pi}\right)^3\,\bigg(
              -{4613165\over 864} 
              + 1340\,\zeta_2 
              + {121075\over 36}\,\zeta_3 
              - 1270\,\zeta_5 
\nonumber\\&&\mbox{\hspace{1em}}
              + n_f\,\Big(
                  {72197\over 162} 
                  - {209\over 2}\,\zeta_2 
                  - {656\over 3}\,\zeta_3 
                  + 5\,\zeta_4 
                  + 55\,\zeta_5
                  \Big)
\nonumber\\&&\mbox{\hspace{1em}}
              + n_f^2\,\Big(
                  -{13171\over 1944} 
                  + {16\over 9}\,\zeta_2 
                  + {26\over 9}\,\zeta_3
                  \Big) 
              \bigg)
          \bigg]\\
          &\approx&
    {m_Q^2\over s}\,\bigg[
      -6 
      - 22\,{\alpha_s\over \pi} 
      + \left({\alpha_s\over \pi}\right)^2\, \Big(
          -108.140
          + 4.48524\,n_f
          \Big)
\nnb\\&&\mbox{}
      + \left({\alpha_s\over \pi}\right)^3\,\Big(
          -409.247
          + 73.3578\,n_f 
          - 0.378270\,n_f^2
          \Big)
\bigg]\,.
\end{eqnarray}
%

The quadratic mass terms in $r_q$ contribute in $\order{\alpha_s^3}$ and
higher only~\cite{CheKue90}.
Finally, we present the quartic mass corrections up to $\order{\alpha_s^3}$:
%
\begin{eqnarray}
r_{q,4} &=&
      \left({m_Q^2\over s}\right)^2\,\bigg[
          \left({\alpha_s\over \pi}\right)^2\,\bigg(
              {13\over 3} 
              - \logmsms 
              - 4\,\zeta_3
              \bigg) 
\nonumber\\&&\mbox{\hspace{0em}}
          + \left({\alpha_s\over \pi}\right)^3\,\bigg(
              -{9707\over 144} 
              + 2\,\logmsms^2 
              + 15\,\zeta_2 
              + 25\,\zeta_3 
              + {50\over 3}\,\zeta_5
              + \logmsms\,\Big(
                  {43\over 12} 
                  - 22\,\zeta_3
                  \Big) 
\nonumber\\&&\mbox{\hspace{1em}}
              + n_f\,\Big(
                  {457\over 108} 
                  - {2\over 3}\,\zeta_2 
                  - {22\over 9}\,\zeta_3 
                  + \logmsms\,(
                      -{13\over 18} 
                      + {4\over 3}\,\zeta_3
                      )
                  \Big) 
              \bigg)
          \bigg]\\
 &\approx&
      \left({m_Q^2\over s}\right)^2\,\bigg[
          \left({\alpha_s\over \pi}\right)^2\,(
              -0.474894
              - \logmsms
              ) 
\nonumber\\&&\mbox{\hspace{1em}}
          + \left({\alpha_s\over \pi}\right)^3\,\Big(
            4.59784 
            - 22.8619\,\logmsms 
            + 2\,\logmsms^2  
\nnb\\&&\mbox{\hspace{2em}}
            + (0.196497 + 0.88052\,\logmsms)\,n_f
              \Big)
          \bigg]\,,
          \nonumber\\[1em]
  r_{Q,4}^{(v)} &=&
      \left({m_Q^2\over s}\right)^2\,\bigg[
          -6 
          - 22\,{\alpha_s\over \pi} 
          + \left({\alpha_s\over \pi}\right)^2\,\bigg(
              -{2977\over 12} 
              + 162\,\zeta_2 
              + 108\,\zeta_3
              - {13\over 2}\,\logmsms 
\nonumber\\&&\mbox{\hspace{1em}}
              + n_f\,\Big(
                  {143\over 18} 
                  + {1\over 3}\,\logmsms 
                  - 4\,\zeta_2 
                  - {8\over 3}\,\zeta_3
                  \Big) 
              \bigg) 
\nonumber\\&&\mbox{\hspace{0em}}
          + \left({\alpha_s\over \pi}\right)^3\,\bigg(
              -{1264093\over 432} 
              + {12099\over 4}\,\zeta_2 
              + {64123\over 18}\,\zeta_3 
              - {13285\over 9}\,\zeta_5 
\nonumber\\&&\mbox{\hspace{1em}}
              + \logmsms\,\Big(
                  -{1309\over 6} 
                  - 22\,\zeta_3
                  \Big) 
              + 13\,\logmsms^2 
              + n_f\,\Big(
                  {130009\over 648} 
                  - {574\over 3}\,\zeta_2 
                  - {1672\over 9}\,\zeta_3 
\nonumber\\&&\mbox{\hspace{2em}}
                  + 10\,\zeta_4 
                  + {440\over 9}\,\zeta_5
                  + \logmsms\,(
                      {199\over 12} 
                      + {4\over 3}\,\zeta_3
                      ) 
                  - {2\over 3}\,\logmsms^2 
                  \Big)
\nonumber\\&&\mbox{\hspace{1em}}
              + n_f^2\,\Big(
                  -{463\over 972} 
                  + {23\over 9}\,\zeta_2 
                  + {28\over 27}\,\zeta_3
                  - {5\over 27}\,\logmsms 
                  \Big) 
              \bigg)
          \bigg]\\
          &\approx&
 \left({m_Q^2\over s}\right)^2\,\bigg[-6 - 22\,{\alpha_s\over \pi}
\nnb\\&&\mbox{\hspace{0em}}
 + \left({\alpha_s\over \pi}\right)^2\,(
     148.218 
     - 6.5\,\logmsms 
     + (-1.84078 + 0.333333\,\logmsms)\,n_f
)
 \nonumber\\&&\mbox{\hspace{0em}}
  + \left({\alpha_s\over \pi}\right)^3\,(
     4800.95 
     - 244.612\,\logmsms + 13\,\logmsms^2
 \nonumber\\&&\mbox{\hspace{1em}}
     + (-275.898 
        + 18.1861\,\logmsms 
        - 0.666667\,\logmsms^2 )\,n_f
 \nonumber\\&&\mbox{\hspace{1em}}
     + (4.97396 
        - 0.185185\,\logmsms)\,n_f^2
\bigg]\,,
          \nonumber\\[1em]
 r_{Q,4}^{(a)} &=&
      \left({m_Q^2\over s}\right)^2\,\bigg[
          6 
          + 10\,{\alpha_s\over \pi} 
          + \left({\alpha_s\over \pi}\right)^2\,\bigg(
              {1147\over 4} 
              - 162\,\zeta_2 
              - 224\,\zeta_3 
              + {75\over 2}\,\logmsms 
\nonumber\\&&\mbox{\hspace{1em}}
              + n_f\,\Big(
                  -{41\over 6} 
                  + 4\,\zeta_2 
                  + {16\over 3}\,\zeta_3
                  - {7\over 3}\,\logmsms 
                  \Big)
              \bigg) 
\nonumber\\&&\mbox{\hspace{0em}}
          + \left({\alpha_s\over \pi}\right)^3\,\bigg(
              {221269\over 48} 
              - {10881\over 4}\,\zeta_2 
              - {20147\over 6}\,\zeta_3 
              - {2225\over 3}\,\zeta_5 
\nonumber\\&&\mbox{\hspace{1em}}
              + \logmsms\,\Big(
                  {4385\over 6} 
                  - 22\,\zeta_3
                  \Big) 
              - 75\,\logmsms^2 
\nonumber\\&&\mbox{\hspace{1em}}
              + n_f\,\Big(
                  -{200923\over 648} 
                  + 196\,\zeta_2 
                  + {5002\over 27}\,\zeta_3 
                  - 10\,\zeta_4 
                  + {1040\over 27}\,\zeta_5
\nonumber\\&&\mbox{\hspace{2em}}
                  + \logmsms\,(
                      -{2323\over 36} 
                      + {4\over 3}\,\zeta_3
                      ) 
                  + {14\over 3}\,\logmsms^2 
                  \Big)
\nonumber\\&&\mbox{\hspace{1em}}
              + n_f^2\,\Big(
                  {2995\over 972} 
                  - {29\over 9}\,\zeta_2 
                  - {20\over 27}\,\zeta_3
                  + {23\over 27}\,\logmsms 
                  \Big) 
              \bigg)
          \bigg]
  \label{eq::r4}
      \\
&\approx&
\left({m_Q^2\over s}\right)^2\,\bigg[
  6 + 10\,{\alpha_s\over \pi} 
\nonumber\\&&\mbox{\hspace{0em}}
  + \left({\alpha_s\over \pi}\right)^2\,
  \Big(
     -248.99 
     + 37.5\,\logmsms 
     + (6.15737 - 2.33333\,\logmsms)\,n_f
\Big)
\nonumber\\&&\mbox{\hspace{0em}}
+ \left({\alpha_s\over \pi}\right)^3\,
  \Big(
     -4670.22 
     + 704.388\,\logmsms 
     - 75\,\logmsms^2 
\nonumber\\&&\mbox{\hspace{1em}}
     + (264.151 - 62.9250\,\logmsms + 4.66667\,\logmsms^2 )\,n_f
\nonumber\\&&\mbox{\hspace{1em}}
     + (-3.10948 + 0.851852\,\logmsms)\,n_f^2
\Big)\bigg]\,.
  \nnb
    \end{eqnarray}
%

The singlet contributions in the vector case are numerically small:
%
\begin{eqnarray}
r^{(v)}_{\rm sing} &=& \left({\alpha_s\over \pi}\right)^3\,\Bigg\{
\left( v_Q + \sum_{q} v_q\right)^2\,
\bigg({55\over 216} - {5\over 9}\zeta_3\bigg)
\nonumber\\&&\mbox{}
+\left({m_Q^2\over s}\right)^2\,v_Q\,\left( v_Q + \sum_{q} v_q\right)
\,\bigg(-{20\over 9} + {50\over 3}\,\zeta_3\bigg)\Bigg\}
\nonumber\\
&\approx&
\left({\alpha_s\over \pi}\right)^3\,\Bigg\{
-0.413180\,\left(v_Q + \sum_q v_q\right)^2\, 
\nonumber\\&&\mbox{\hspace{1em}}
+ 17.8121\,v_Q\,\left( v_Q + \sum_{q} v_q\right)\,
        \left({m_Q^2\over s}\right)^2\Bigg\} + 
 \order{\left({m_Q^2\over s}\right)^3}\,.
\label{eq::Rvs}
\end{eqnarray}
%
 
They do not receive $m^2$ corrections~\cite{CheKue90}.

The singlet contributions in the axial-vector case are exceptional in
the sense that for bottom quark production the top quark does not
decouple, i.e., the contributions where the top quark couples to one of
the external currents are not suppressed by powers of $1/m_t^2$.  The
corresponding corrections up to $\order{m_b^2\alpha_s^3}$ have been
computed in~\cite{KniKue90,CheKue93,CheTar94LarRitVer94,CheKwi93}.  They
turn out to be small, so we will neglect them in the following.

Considering the axial vector singlet contribution to top quark
production, on the other hand, one should take into account the
full top-bottom doublet, because the axial anomaly cancels in this
combination. However, this strategy induces completely massless final
states without any top quarks. 
At order $\alpha_s^2$ the expansion for the contributions from the full
top-bottom doublet starts with the $m^6$ terms \cite{HarSte:eett}. The
separate piece from massless cuts is known in analytical form
\cite{KniKue90} and has to be subtracted if one wants to obtain the
contribution from the final states with top quarks. The complete result
is given in Ref.~\cite{HarSte:eett}. Thus the singlet contribution to
$t\bar t$ production is given in expanded form by
\begin{eqnarray}
  r_{\rm sing} &=& {1\over 4} \left({\alpha_s\over \pi}\right)^2\,\Bigg\{
      {5\over 4} 
      - {2\over 3}\,\zeta_2 
      + {m^2\over s}\,\bigg(
          - {4\over 3} 
          + {20\over 3}\,\zeta_2
          + 4\,\logmsms 
          - {2\over 3}\,\logmsms^2 
          \bigg) 
\nonumber\\  && 
      + \left({m^2\over s}\right)^2\,\bigg[
          {13\over 2} 
          + {2\over 3}\,\zeta_2 
          - {16\over 3}\,\zeta_3
          + \logmsms\,\Big(
              - 5 
              + {4\over 3}\,\zeta_2
              \Big) 
          - {1\over 3}\,\logmsms^2 
          - {2\over 9}\,\logmsms^3 
          \bigg]
      \Bigg\}\nnb\\
&& +\,\order{\alpha_s^3}\nnb\\
&=&
{1\over 4} \left({\alpha_s\over \pi}\right)^2
 \bigg[0.153377 + {m^2\over s}\,(9.63289 + 4\, \logmsms -
0.666667\, \logmsms^2 )\nnb\\&& + 
 \left({m^2\over s}\right)^2\,  (1.18565 - 2.80675\, \logmsms -
 0.333333\, \logmsms^2  - 0.222222\, \logmsms^3 )\bigg]
\nnb\\&& +\,\order{\alpha_s^3}\,,
\end{eqnarray}
where the weak coupling $a_t^2=a_b^2=-a_t a_b=1/4$ 
has been pulled out for clarity.
The separate contribution from the massless cuts is not yet available in
order $\alpha_s^3$. In principle it should be subtracted from the
following result in order to 
arrive at the production rate for top quarks:
\begin{eqnarray}
r'_{\rm sing} &=&
 {1\over 4}\,
        \left({\alpha_s\over \pi}\right)^3\,\left({m_t^2\over s}\right)^2\,
   \bigg[-{380\over 3}\,\zeta_3 + {1520\over 3}\,\zeta_5
     + n_f\,\Big({40\over 9}\,\zeta_3 - {160\over
        9}\,\zeta_5\Big)\bigg]\,
\nonumber\\&\approx& {1\over 4}\,
        \left({\alpha_s\over \pi}\right)^3\,
        \left({m_t^2\over s}\right)^2\left(373.116 - 13.0918\, n_f \right)\,.
\end{eqnarray}
The prime indicates that this expression still contains the
contributions from purely massless final states, as mentioned before. We
explicitly denote the mass by $m_t$ here in order to recall that, on
the one hand, this result applies only to top production, and on the
other hand, the full $(t,b)$ doublet has been taken into account.

Let us now investigate the numerical significance of mass effects for
the charm and bottom case.  If not stated otherwise, we will adopt the
following input data:
\begin{eqnarray}
&& \hspace{0em}
M_Z = \nzmass\ {\rm GeV}\,,\qquad \alpha_s(M_Z^2) =
\nalphas\,,\nonumber\\
&&
m_c(m_c) = \nmcmc\ {\rm GeV}\,, \qquad 
m_b(m_b) = \nmbmb\ {\rm GeV}\,,\qquad
m_t(m_t) = \nmtmt\,{\rm GeV}\,.
\label{eq::input}
\end{eqnarray}
The running of the quark masses and the coupling constant to the scale
$\sqrt{s}$ is performed with three loop accuracy. Since we are working in
the $\msbar$ scheme, we have to take into account the matching
conditions for $\alpha_s^{(n_f)}$ when going from $n_f$ to $n_f\pm 1$.
We perform the matching from $n_f=5$ to $n_f=4$ at $m_b(m_b)$ and from
$n_f=5$ to $n_f=6$ at $m_t(m_t)$. For illustration, some values for
$\alpha_s(s)$, $m_c(\sqrt{s})$, $m_b(\sqrt{s})$, and $m_t(\sqrt{s})$ are
displayed in Table~\ref{tab::runam}. The charm, bottom, and top masses
are defined for $n_f=4$, $5$, and $6$, respectively.
As stated above, the approximation is expected to become unreliable in
the threshold region, that means below $5.5-6$~GeV for charm,
$12-12.5$~GeV for bottom, and 
$420-450$~GeV for top contributions, which justifies the choice for the
cms energies in the figures presented below.

In Figs.~\ref{fig::r0}--\ref{fig::rQ4} we display separately the
contributions for the massless case (Fig.~\ref{fig::r0}) and for the
$m^2$ and $m^4$ terms, including successively higher orders in
$\alpha_s$.  While Fig.~\ref{fig::rq} shows the effects of the diagrams
with light quarks coupling to the external current, Figs.~\ref{fig::rQ2}
and \ref{fig::rQ4} correspond to the case where the massive quark
couples directly to the external current.   The axial contribution is
presented for the top quark only.

The variation of the prediction with the renormalization scale $\mu$ is
shown in Figs.~\ref{fig::mudep0} to \ref{fig::mudep4}.


\begin{table}
\begin{center}
\begin{tabular}{l}
\begin{tabular}{|l|lllllll|}
\hline $ \sqrt{s} $ & $ 5. $ & $ 6. $ & $ 7. $ & $ 8. $ & $ 9. $ & $
10. $ & $ 10.5 $ \\\hline $ \alpha_s^{(4)}(s) $ & $ 0.212 $ & $ 0.2 $ &
$ 0.192 $ & $ 0.185 $ & $ 0.179 $ & $ 0.174 $ & $ 0.172 $ \\ $
m^{(4)}_c(\sqrt{s}) $ & $ 0.829 $ & $ 0.803 $ & $ 0.784 $ & $ 0.769 $ &
$ 0.756 $ & $ 0.745 $ & $ 0.74 $ \\ \hline
\end{tabular}\\[3em]
\begin{tabular}{|l|lllll|}
\hline $ \sqrt{s} $ & $ 11. $ & $ 12. $ & $ 14. $ & $ 16. $ & $ 20. $
 \\\hline $ \alpha_s^{(5)}(s) $ & $ 0.174 $ & $ 0.171 $ & $ 0.165 $ & $
 0.161 $ & $ 0.153 $ \\ $ m^{(5)}_b(\sqrt{s}) $ & $ 3.61 $ & $ 3.57 $ &
 $ 3.5 $ & $ 3.44 $ & $ 3.35 $ \\ \hline
\end{tabular}\\[3em]
\begin{tabular}{|l|lll|}
\hline $ \sqrt{s} $ & $ 420. $ & $ 460. $ & $ 500. $ \\\hline $
 \alpha_s^{(6)}(s) $ & $ 0.097 $ & $ 0.0961 $ & $ 0.0952 $ \\ $
 m^{(6)}_t(\sqrt{s}) $ & $ 154. $ & $ 153. $ & $ 152. $ \\ \hline
\end{tabular}
\end{tabular}
\parbox{\captionwidth}{ \caption[]{\label{tab::runam} Running coupling
and masses at different scales.  Matching for $\alpha_s$ is performed at
the values $m(m)$ given in (\ref{eq::input}).  }}
\end{center}
\end{table}


\begin{table}
\begin{center}
\begin{tabular}{|l|lllll|}
\hline & $ \alpha_s^0 $ & $ \alpha_s^1 $ & $ \alpha_s^2 $ & $ \alpha_s^3
 $ & $ \Sigma $ \\\hline $ m^0 $ & $ 1. $ & $ 0.05482 $ & $ 0.004581 $ &
 $ -0.001898 $ & $ 1.058 $ \\ $ m^2 $ & $ 0 $ & $ 0.003264 $ & $
 0.001628 $ & $ 0.000519 $ & $ 1.063 $ \\ $ m^4 $ & $ -0.0001477 $ & $
 -0.00002969 $ & $ 0.00001245 $ & $ 0.00002026 $ & $ 1.063 $ \\ \hline
\end{tabular}
\parbox{\captionwidth}{
  \caption[]{$r^{(v)}_Q$ for $\sqrt{s}=10.5\ {\rm GeV}$.}}
\end{center}
\end{table}

As discussed already in earlier papers, the size of the higher order
corrections decreases quickly with increasing order in $\alpha_s$, as
far as the massless approximation is concerned
(Fig.~\ref{fig::r0}). This is reflected in the stability of the result
under variation of the renormalization scale $\mu$ by a factor between
1/2 and 2 as displayed in Fig.~\ref{fig::mudep0}. The $m^2$ terms in
$r_q$ (Fig.~\ref{fig::rq}~(a), (c), (e)) arise for the first time in
order $\alpha_s^3$. As a consequence this prediction exhibits a strong
$\mu$ dependence as displayed in Fig.~\ref{fig::mudepv0}~(a), (c), and
(e). However, this term is typically around $10^{-4}$ and thus
irrelevant for all practical purposes.

The quartic terms in $r_q$ which are displayed in
Figs.~\ref{fig::rq}~(b), (d), and (f) contribute in
$\order{\alpha_s^2}$ and higher. Terms of order $m^4\alpha_s^3$ and
$m^4\alpha_s^2$ are of comparable magnitude, not only for the low energy
case but even at the highest energy, whence the prediction for the $m^4$
terms in $r_q$ must be considered as uncertain within a factor two. This
is reflected in the strong $\mu$ dependence of the 
result shown in Fig.~\ref{fig::mudepv0}~(b), (d), and (e).

No reduction of this variation is observed when moving from the
$\alpha_s^2$ to the $\alpha_s^3$ calculation. Nevertheless, the
prediction for the total cross section is not seriously affected by
this instability  since these terms are again of order $10^{-4}$ only
and thus far below the foreseeable experimental precision.

The dominant quartic mass terms are obviously expected from $r_Q$, the
part where the massive quark is coupled to the external current. For
comparison we also discuss the quadratic terms which are known since
long and are shown in Fig.~\ref{fig::rQ2}. The vector current induced
piece is displayed in Fig.~\ref{fig::rQ2} for $c$, $b$, and $t$ quarks,
the axial piece is shown for $t$ quarks only. Terms of increasing order
in $\alpha_s$ decrease in magnitude and apparent convergence is
observed. This welcome behavior is reflected by the improved stability
of higher orders under variations of $\mu$ (Fig.~\ref{fig::mudep2}). 
The behavior of the quartic
terms which are the theme of this paper is more problematic
(Fig.~\ref{fig::rQ4}). The bulk of the result is given by the Born
approximation. The order $\alpha_s$-, $\alpha_s^2$-, and
$\alpha_s^3$-corrections are significantly smaller than the leading
terms. However, with increasing order they do  not decrease but remain 
roughly comparable in magnitude. This is reflected in the strong $\mu$
dependence displayed in Fig.~\ref{fig::mudep4}. Depending on the choice
of $\mu$, the relative size of the three corrections varies drastically.

Nevertheless, the higher orders are small compared to the $m^4$ Born
terms and, even more important, their instability does not affect the
stability of the overall prediction for $R$ resulting from the sum of the
different terms. Accepting as an estimate of the uncertainty of
$r_{Q,4}$ its variation with $\mu$ between $\sqrt{s}/2$ and $2\sqrt{s}$,
$r_{c,4}$ at $6$~GeV varies by $\pm 0.0005$, $r_{b,4}$ at $14$~GeV by
$\pm 0.0016$. For $r_{t,4}$ the variation is negligible.

These considerations demonstrate that a prediction for $R$ has been
obtained which is valid in order $\alpha_s^3$ and includes mass terms in
an expansion up to order $m^4$.

From a pragmatic point of view the smallness of the $m^4$ terms, in
particular of their QCD corrections, allows to ignore 
their apparent instability. Nevertheless, one may also attempt to arrive
at a more stable result for the quartic terms by replacing the running
mass by the invariant one ($\hat m$) through
\begin{equation}
m(\mu) = \hat m\,\exp\int\dd a{\gamma_m(a)\over a\beta(a)}\,,
\label{eq::msi}
\end{equation}
where $a \equiv \alpha_s(\mu^2)/\pi$.
$\beta(a)$ and $\gamma_m(a)$ are the renormalization group functions
governing the running of the strong coupling constant and the quark
mass:
\begin{equation}
\mu^2\ddoverdd{\mu^2} a = a\,\beta(a)\,,\qquad
\mu^2\ddoverdd{\mu^2} m = m\,\gamma_m(a)\,.
\end{equation}
Their perturbative expansion is known up to $\order{\alpha_s^4}$ both
for $\beta(a)$ \cite{beta4} and $\gamma_m(a)$ \cite{gammam4}.  The
integral in (\ref{eq::msi}) is solved perturbatively, and the resulting
expression for $R$ is re-expanded in $\alpha_s$ up to the third power.
For the scale invariant mass we assume the values $\hat m_c = 2.8$~GeV,
$\hat m_b = 15.3$~GeV, and $\hat m_t = 1076$~GeV which corresponds to
solving (\ref{eq::msi}) w.r.t.\ $\hat m$ for $\mu=m$ at three-loop order
and using the numerical values of Eq.~(\ref{eq::input}).  The result for
the $\mu$ dependence of $r_{Q,4}^{(v)}$ in this approach is shown in
Fig.~\ref{fig::mudep4si}.

The variation of the ${\cal O}(\alpha_s^3)$ prediction with $\mu$ is
reduced. For example, for $\sqrt{s} = 6$ GeV the ${\cal O}(\alpha_s^3)$
prediction in Fig. 8 varies between $-0.10\cdot 10^{-2}$ and $-0.20\cdot
10^{-2}$ with a central value of $-0.18\cdot 10^{-2}$ at $\mu =
\sqrt{s}$, compared to a range from $-0.13\cdot 10^{-2}$ to $-0.20\cdot
10^{-2}$ with a central value of $-0.19\cdot 10^{-2}$ in Fig.~9. At the
same time one observes a reduction of the higher order terms compared to
the Born approximation and the ${\cal O}(\alpha_s)$ prediction.  For
$\mu = \sqrt{s} $ the results within the two approaches are quite close,
which gives additional confidence in the reliability of these numbers.

\subsubsection*{Summary}

The total cross section for the production of massive quarks in electron
positron annihilation can be predicted in perturbative QCD. After
expansion in $m^2/s$ the quartic terms, i.e. those proportional to
$m^4/s^2$, were calculated up to order $\alpha_s^3$ for vector and axial
current induced rates. Predictions relevant for charm, bottom and top
quark production were presented. The $\alpha_s^3$ corrections were shown
to be comparable to terms of order $\alpha_s$ and $\alpha_s^2$. As a
consequence, the predictions exhibit a sizeable dependence on the
renormalization scale. Adopting instead of the $\msbar$ scheme a
framework where the running mass $\bar{m}(\mu)$ is replaced
by the invariant mass $\hat{m}$, the stability of the prediction is
improved and, at the same time, the relative size of the large order
terms decreases.  Obviously, an improved understanding of the origin of
these large corrections would be highly desirable.

Combining these results with the massless prediction and the quadratic
mass terms we have demonstrated that the cross section for massive
quark production at electron positron colliders is under control in
order $\alpha_s^3$ from the high energy region down to fairly low
energies.

\subsubsection*{Remark} 
The results of this paper are available in
{\tt Mathematica} format at \\
\verb$http://www-ttp.physik.uni-karlsruhe.de/Progdata/ttp00/ttp00-09/$.

\subsubsection*{Acknowledgments}

This work was supported by {\it DFG-Forschergruppe ``Quantenfeldtheorie,
  Computeralgebra und Monte-Carlo-Simulation''}. R.H.\ acknowledges
support by {\it Landesgraduiertenf\"orderung} at the University of
Karlsruhe and by {\it Deutsche Forschungsgemeinschaft}.

\def\app#1#2#3{{\it Act.~Phys.~Pol.~}{\bf B #1} (#2) #3}
\def\apa#1#2#3{{\it Act.~Phys.~Austr.~}{\bf#1} (#2) #3}
\def\cmp#1#2#3{{\it Comm.~Math.~Phys.~}{\bf #1} (#2) #3}
\def\cpc#1#2#3{{\it Comp.~Phys.~Commun.~}{\bf #1} (#2) #3}
\def\epjc#1#2#3{{\it Eur.\ Phys.\ J.\ }{\bf C #1} (#2) #3}
\def\fortp#1#2#3{{\it Fortschr.~Phys.~}{\bf#1} (#2) #3}
\def\ijmpc#1#2#3{{\it Int.~J.~Mod.~Phys.~}{\bf C #1} (#2) #3}
\def\ijmpa#1#2#3{{\it Int.~J.~Mod.~Phys.~}{\bf A #1} (#2) #3}
\def\jcp#1#2#3{{\it J.~Comp.~Phys.~}{\bf #1} (#2) #3}
\def\jetp#1#2#3{{\it JETP~Lett.~}{\bf #1} (#2) #3}
\def\mpl#1#2#3{{\it Mod.~Phys.~Lett.~}{\bf A #1} (#2) #3}
\def\nima#1#2#3{{\it Nucl.~Inst.~Meth.~}{\bf A #1} (#2) #3}
\def\npb#1#2#3{{\it Nucl.~Phys.~}{\bf B #1} (#2) #3}
\def\nca#1#2#3{{\it Nuovo~Cim.~}{\bf #1A} (#2) #3}
\def\plb#1#2#3{{\it Phys.~Lett.~}{\bf B #1} (#2) #3}
\def\prc#1#2#3{{\it Phys.~Reports }{\bf #1} (#2) #3}
\def\prd#1#2#3{{\it Phys.~Rev.~}{\bf D #1} (#2) #3}
\def\pR#1#2#3{{\it Phys.~Rev.~}{\bf #1} (#2) #3}
\def\prl#1#2#3{{\it Phys.~Rev.~Lett.~}{\bf #1} (#2) #3}
\def\pr#1#2#3{{\it Phys.~Reports }{\bf #1} (#2) #3}
\def\ptp#1#2#3{{\it Prog.~Theor.~Phys.~}{\bf #1} (#2) #3}
\def\ppnp#1#2#3{{\it Prog.~Part.~Nucl.~Phys.~}{\bf #1} (#2) #3}
\def\sovnp#1#2#3{{\it Sov.~J.~Nucl.~Phys.~}{\bf #1} (#2) #3}
\def\tmf#1#2#3{{\it Teor.~Mat.~Fiz.~}{\bf #1} (#2) #3}
\def\yadfiz#1#2#3{{\it Yad.~Fiz.~}{\bf #1} (#2) #3}
\def\zpc#1#2#3{{\it Z.~Phys.~}{\bf C #1} (#2) #3}
\def\ibid#1#2#3{{ibid.~}{\bf #1} (#2) #3}

%
\begin{figure}[p]
  \begin{center}
    \leavevmode
    \begin{tabular}{rr}
      (a) & (b) \\[-.5em]
      \epsfxsize=18em
      \epsffile[110 265 465 560]{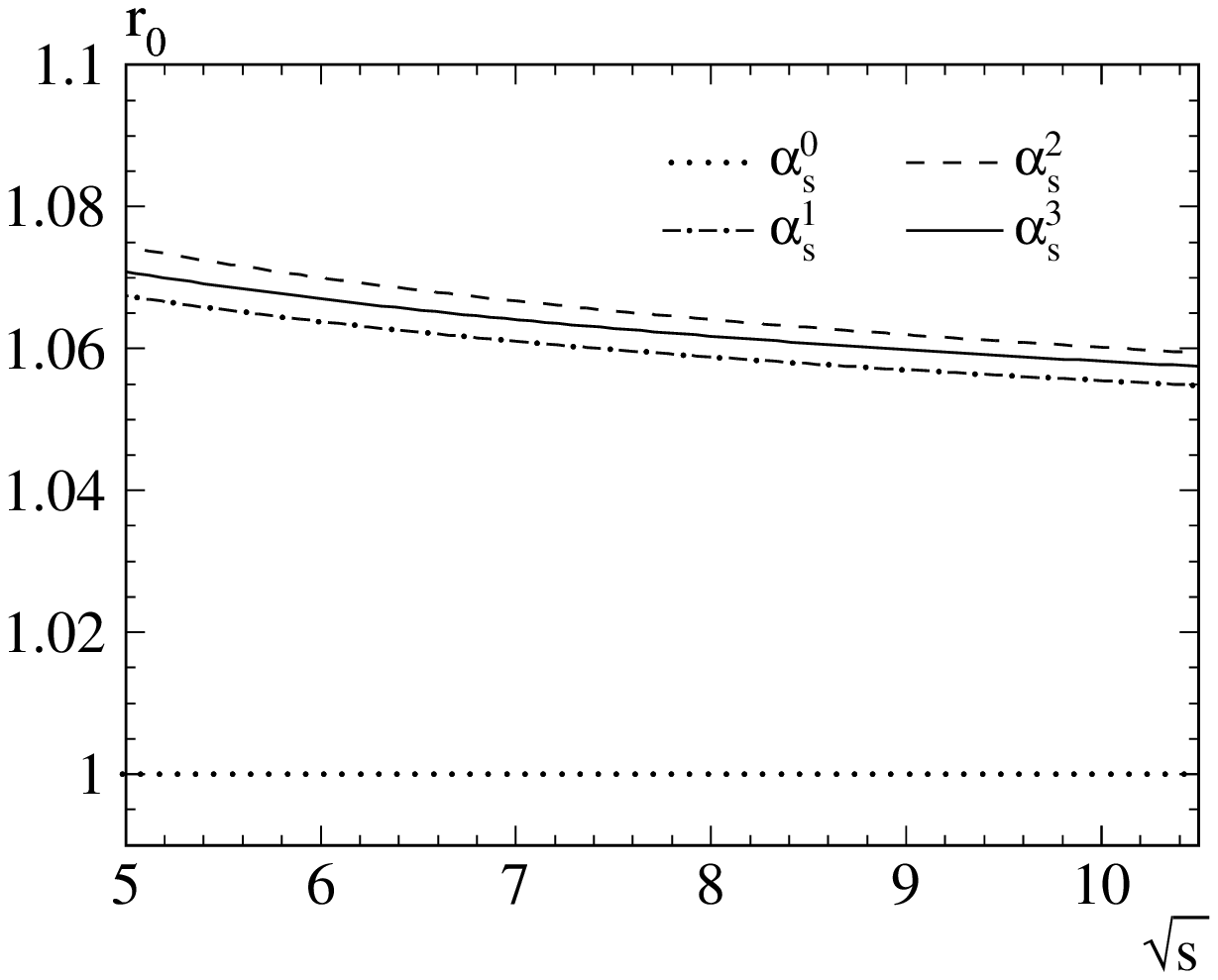} &
      \epsfxsize=18em 
      \epsffile[110 265 465 560]{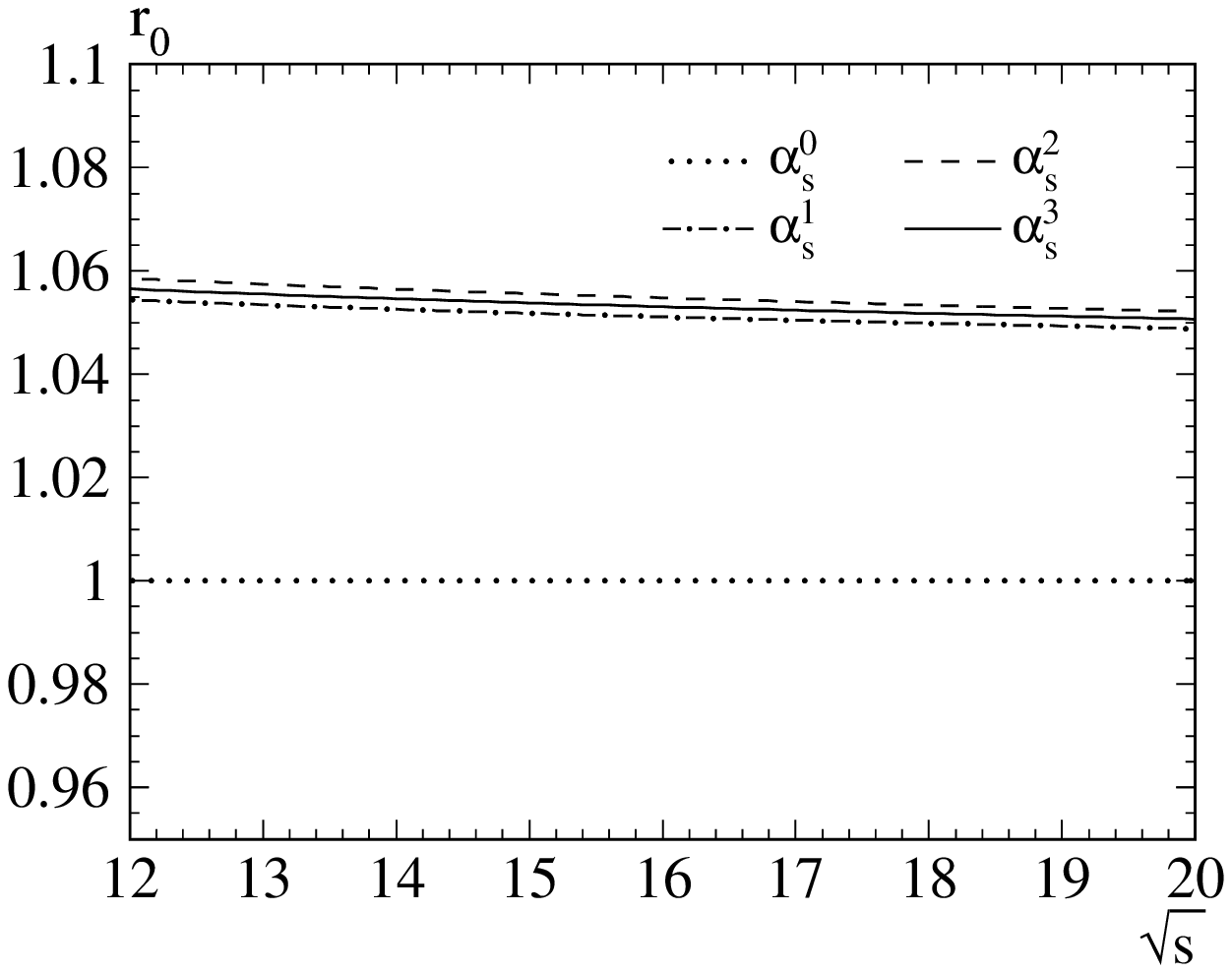} \\[.5em]
       & \multicolumn{1}{c}{\hspace{-2em}(c)}\\[-.5em]
      \multicolumn{2}{c}{
      \epsfxsize=18em
      \epsffile[110 265 465 560]{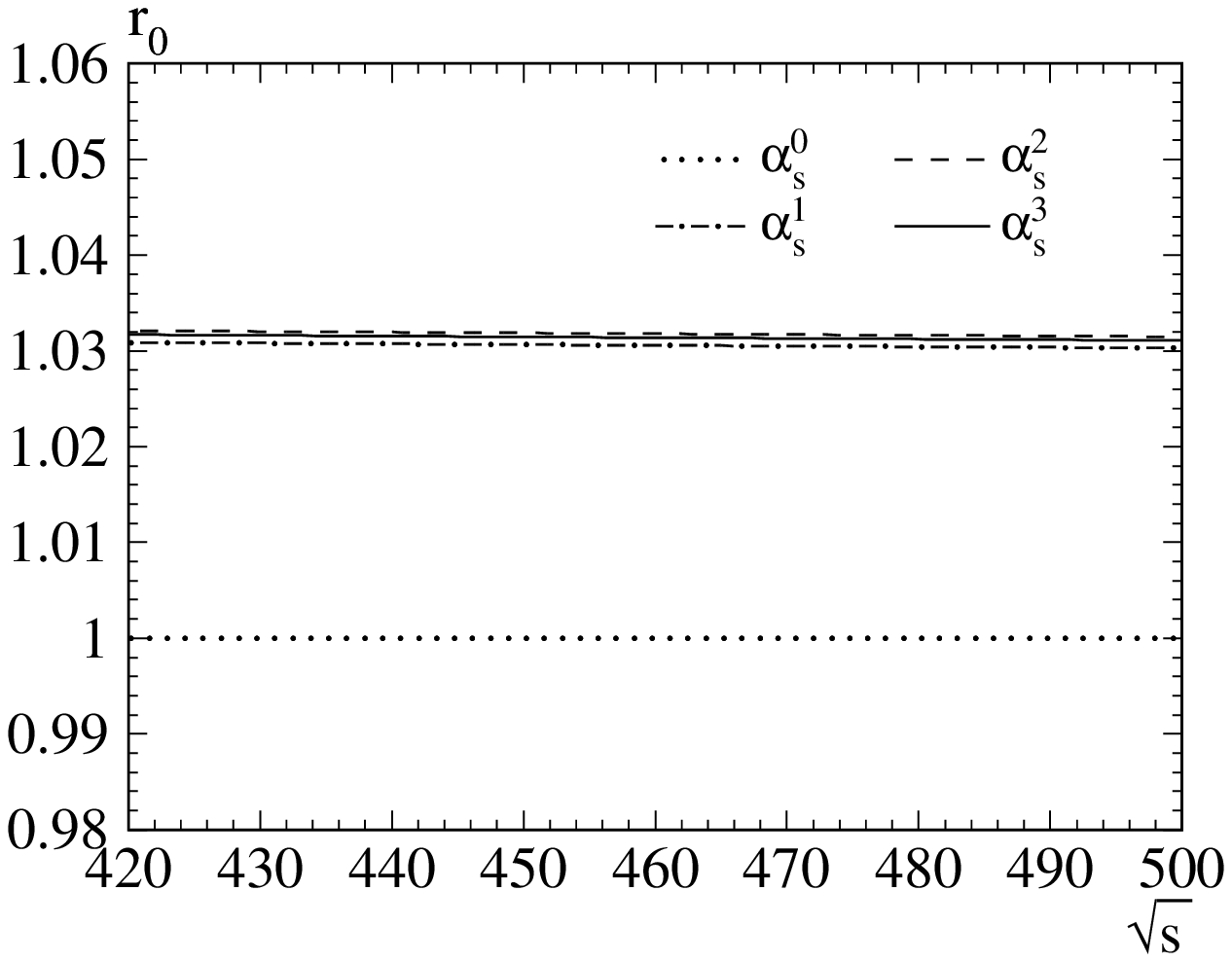}}
    \end{tabular}
    \parbox{\captionwidth}{
      \caption[]{\label{fig::r0}
        The massless approximation $r_0$ in three different energy
        ranges, relevant for (a)~charm, (b)~bottom and (c)~top
        production.
}}
  \end{center}
\end{figure}
%
%
\begin{figure}[p]
  \begin{center}
    \leavevmode
    \begin{tabular}{rr}
      (a) & (b) \\[-.5em]
      \epsfxsize=18em
      \epsffile[110 265 465 560]{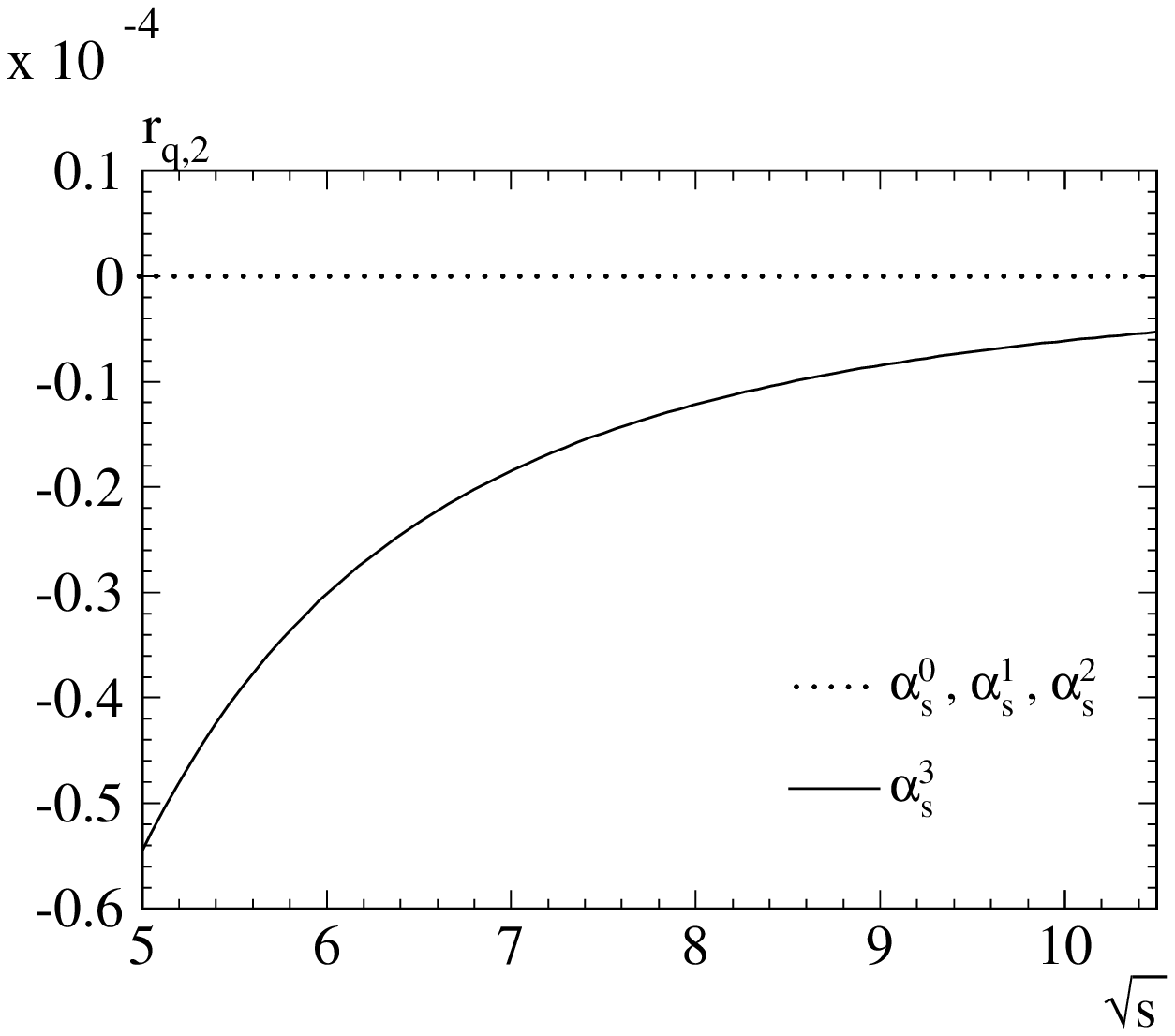} &
      \epsfxsize=18em 
      \epsffile[110 265 465 560]{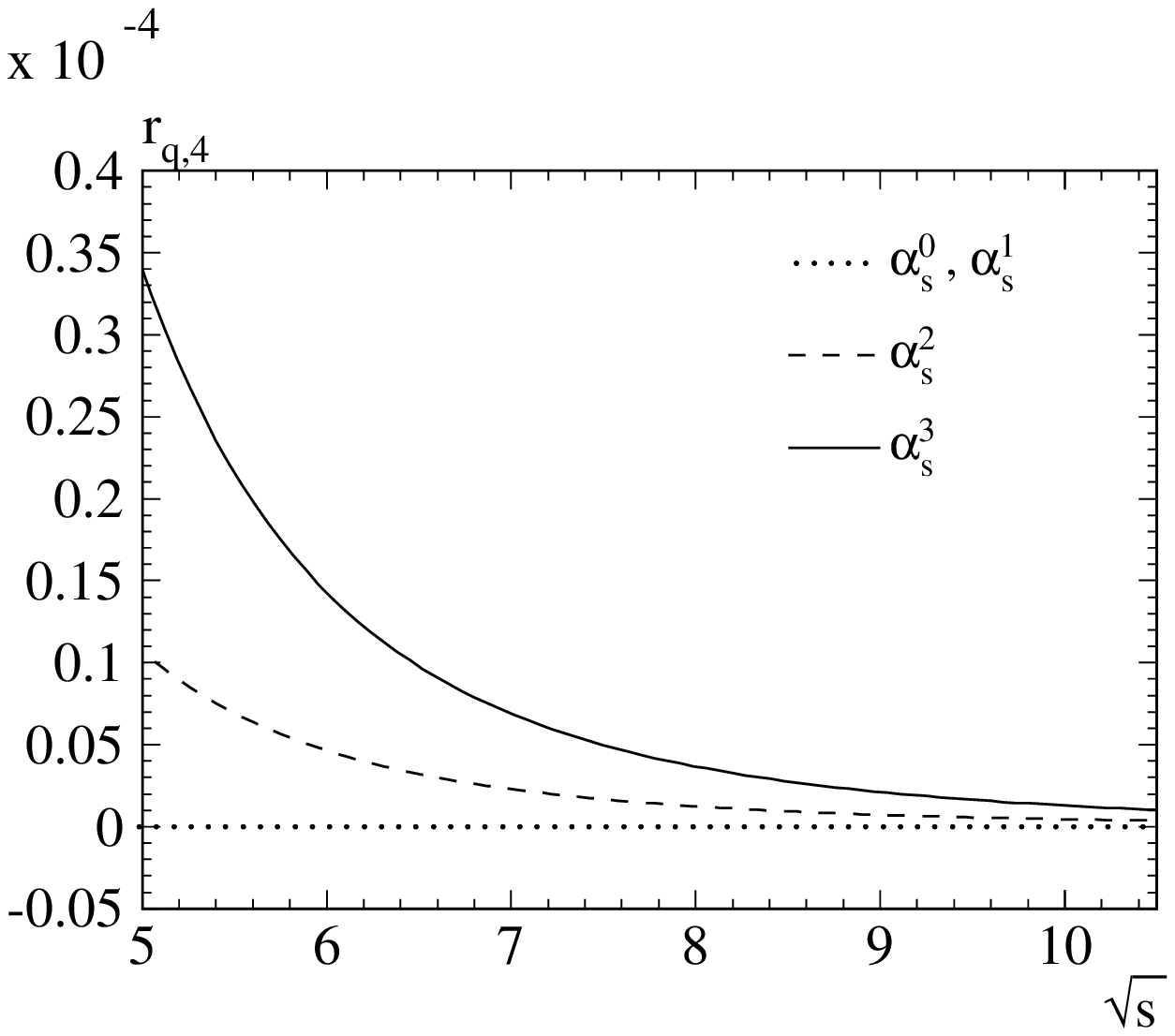} \\[.5em]
      (c) & (d) \\[-.5em]
      \epsfxsize=18em
      \epsffile[110 265 465 560]{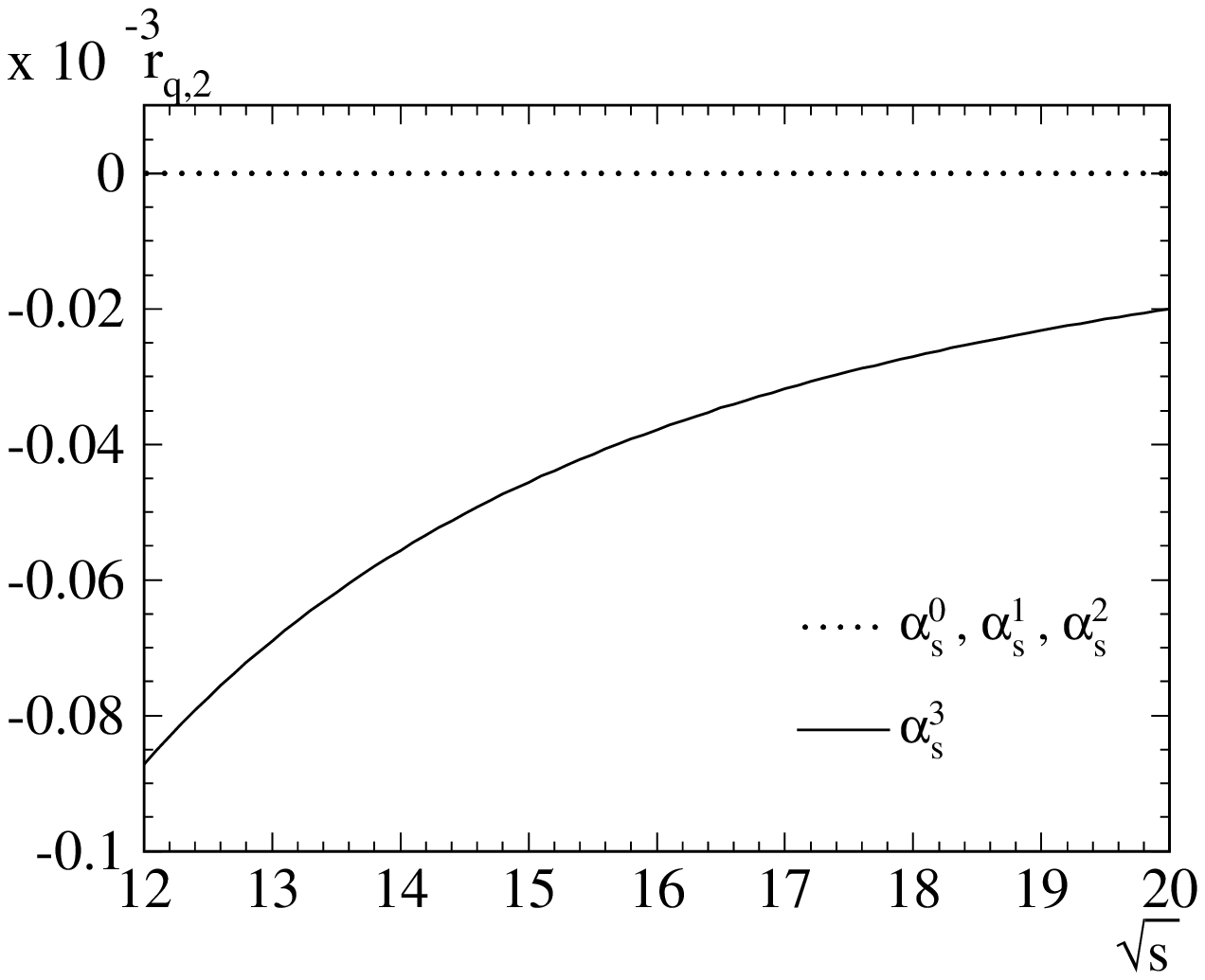} &
      \epsfxsize=18em
      \epsffile[110 265 465 560]{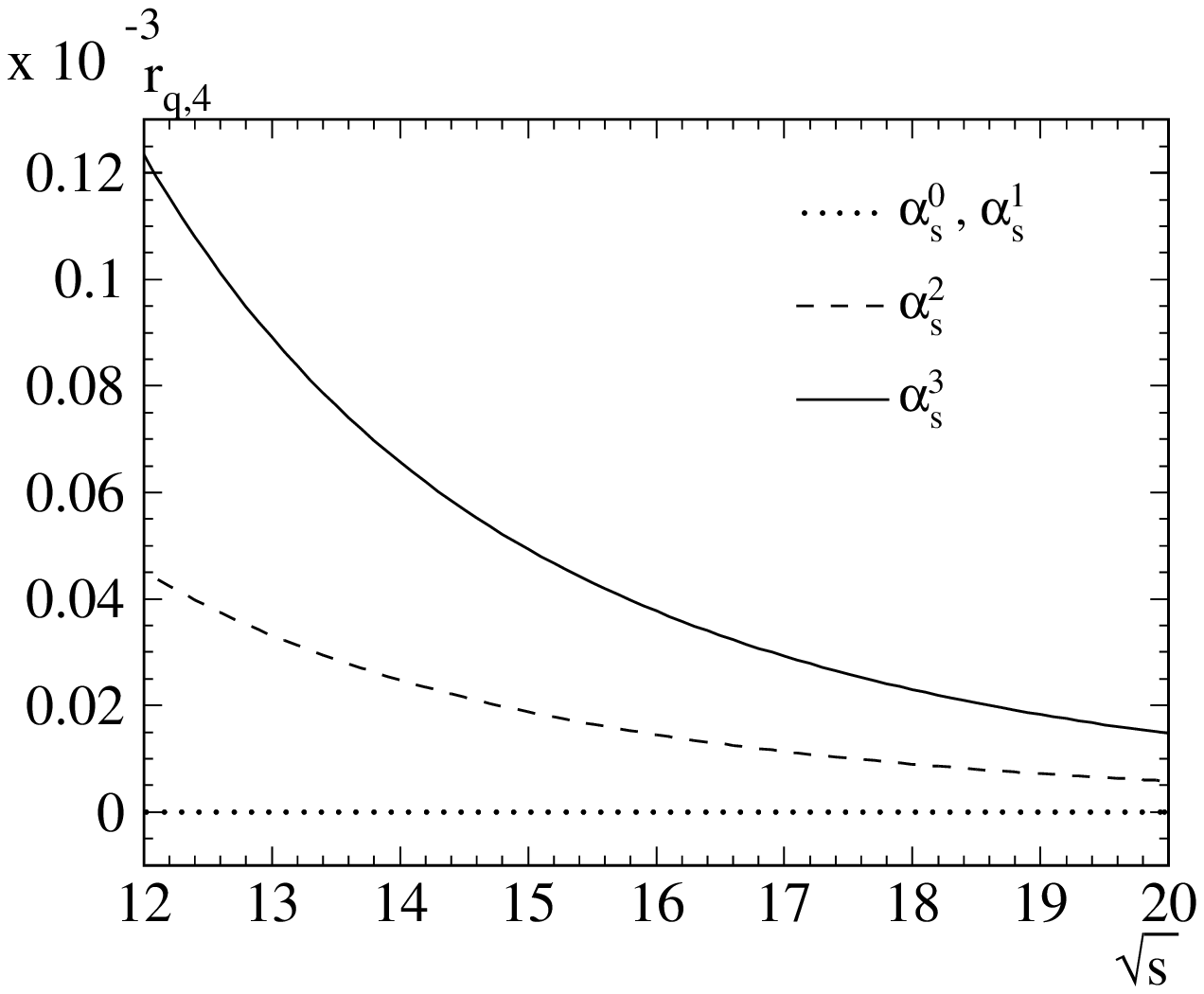} \\[.5em]
      (e) & (f) \\[-.5em]
      \epsfxsize=18em
      \epsffile[110 265 465 560]{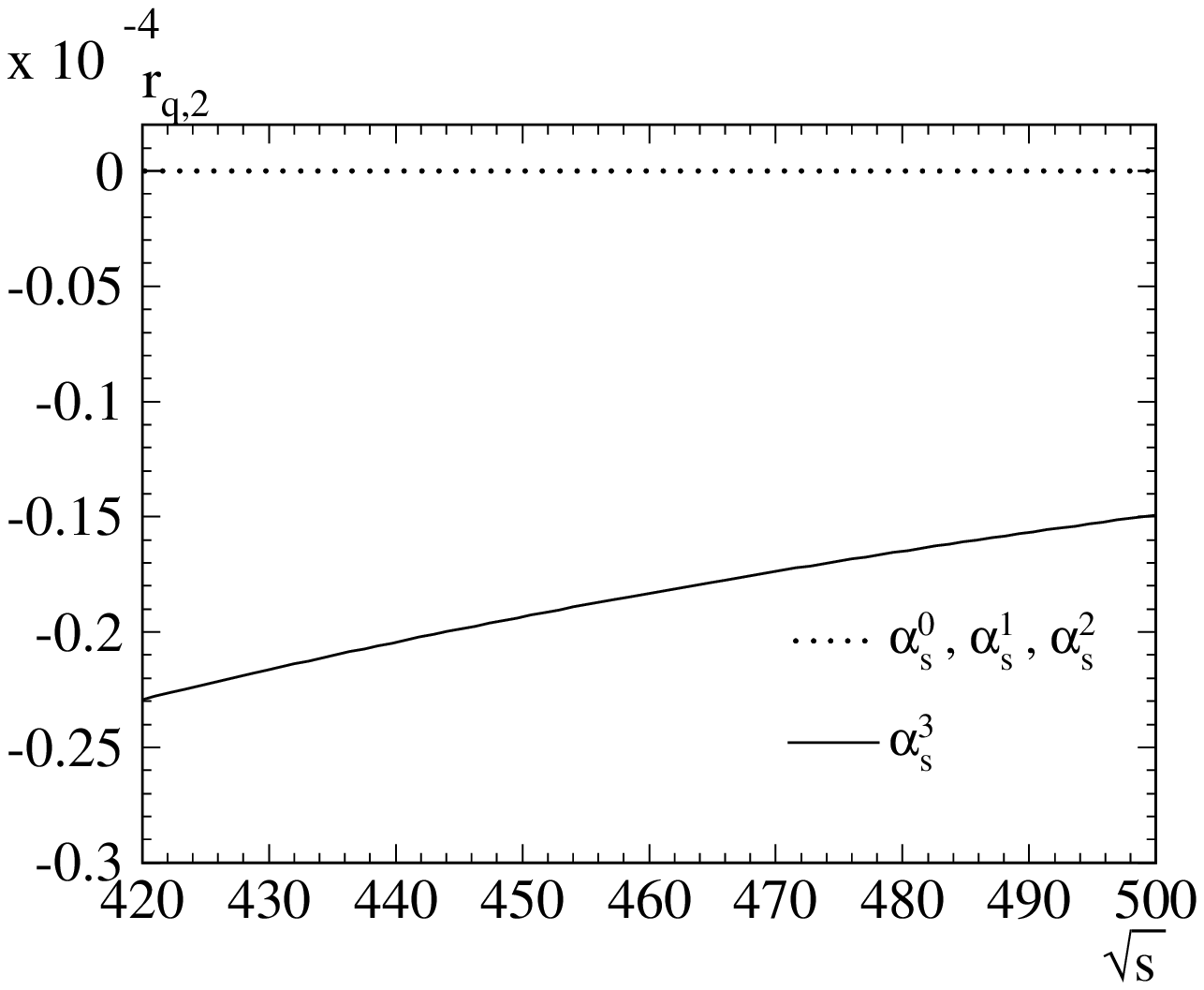} &
      \epsfxsize=18em
      \epsffile[110 265 465 560]{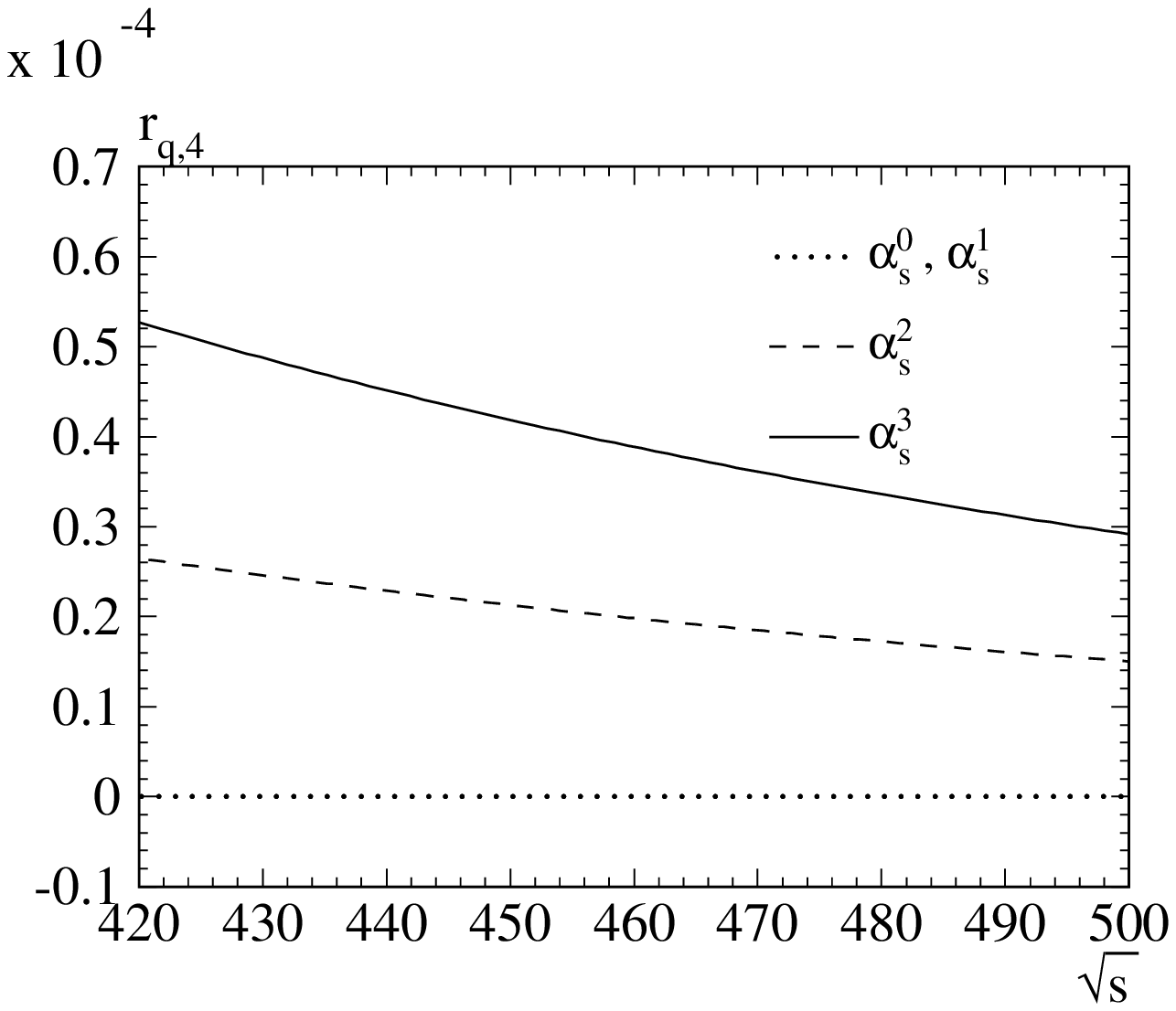}
    \end{tabular}
    \parbox{\captionwidth}{
      \caption[]{\label{fig::rq} 
        Mass corrections to the non-singlet contribution of $r_q$ for
        $c$, $b$, and $t$ production (1.\ ,2., and 3.\ row, respectively), 
        arising from diagrams where the external current
        couples to massless quarks only. Left column: quadratic,
    starting to be non-zero in $\order{\alpha_s^3}$; right
        column: quartic mass corrections, starting to be non-zero in
    $\order{\alpha_s^2}$. 
        }}
  \end{center}
\end{figure}
%
%
\begin{figure}[p]
  \begin{center}
    \leavevmode
    \begin{tabular}{rr}
      (a) & (b) \\[-.5em]
      \epsfxsize=18em
      \epsffile[110 265 465 560]{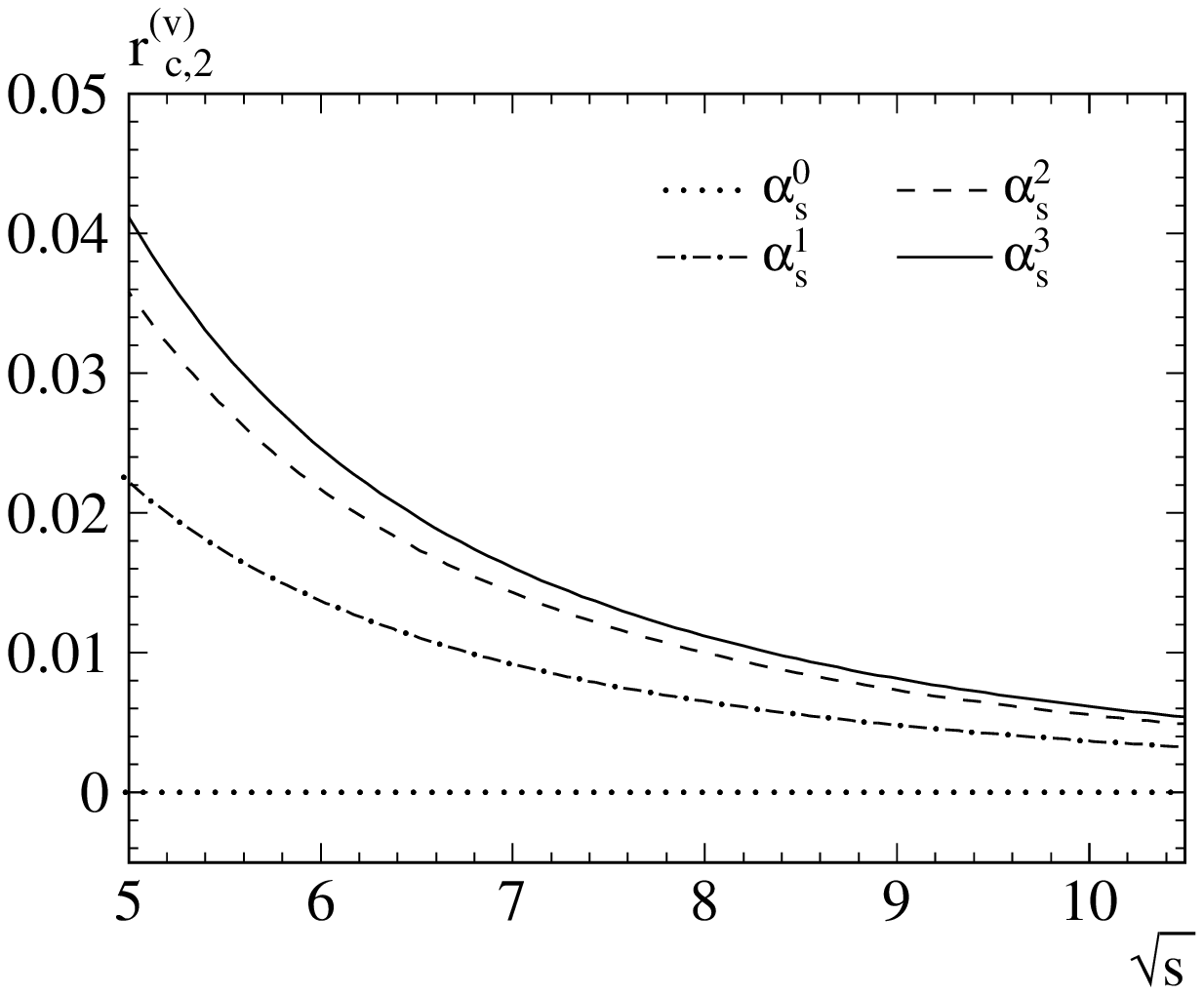} &
      \epsfxsize=18em 
      \epsffile[110 265 465 560]{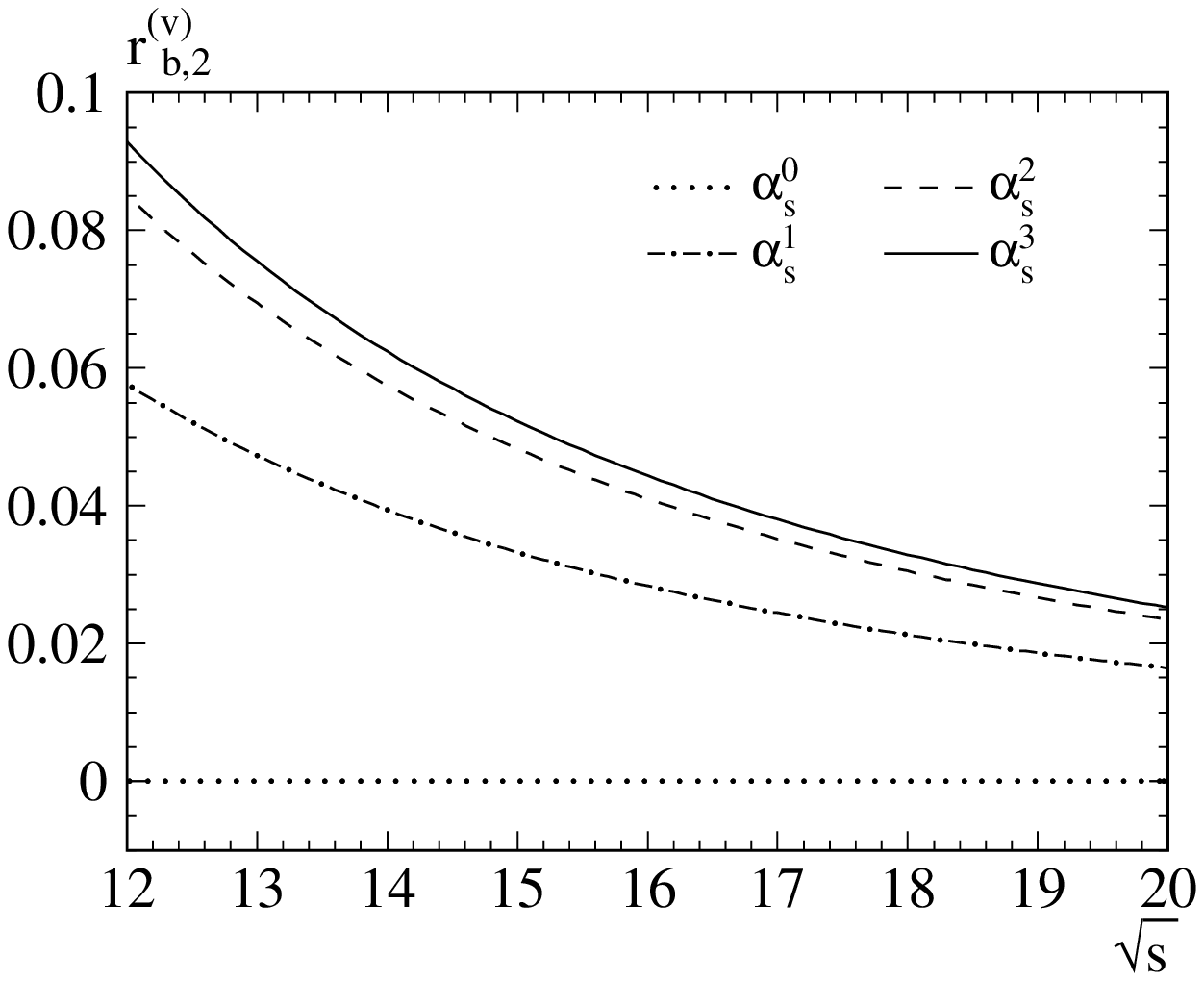} \\[.5em]
      (c) & (d) \\[-.5em]
      \epsfxsize=18em
      \epsffile[110 265 465 560]{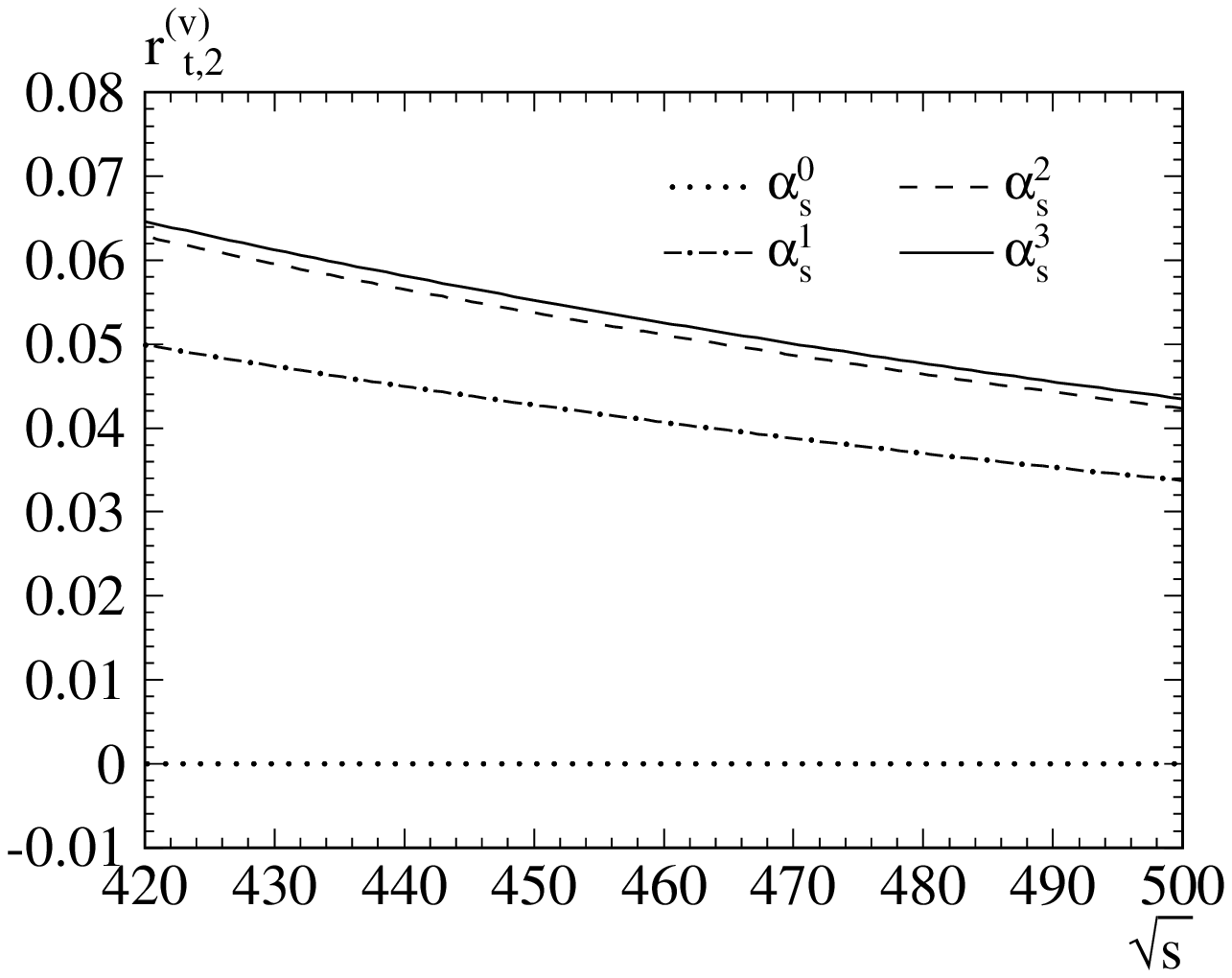} &
      \epsfxsize=18em
      \epsffile[110 265 465 560]{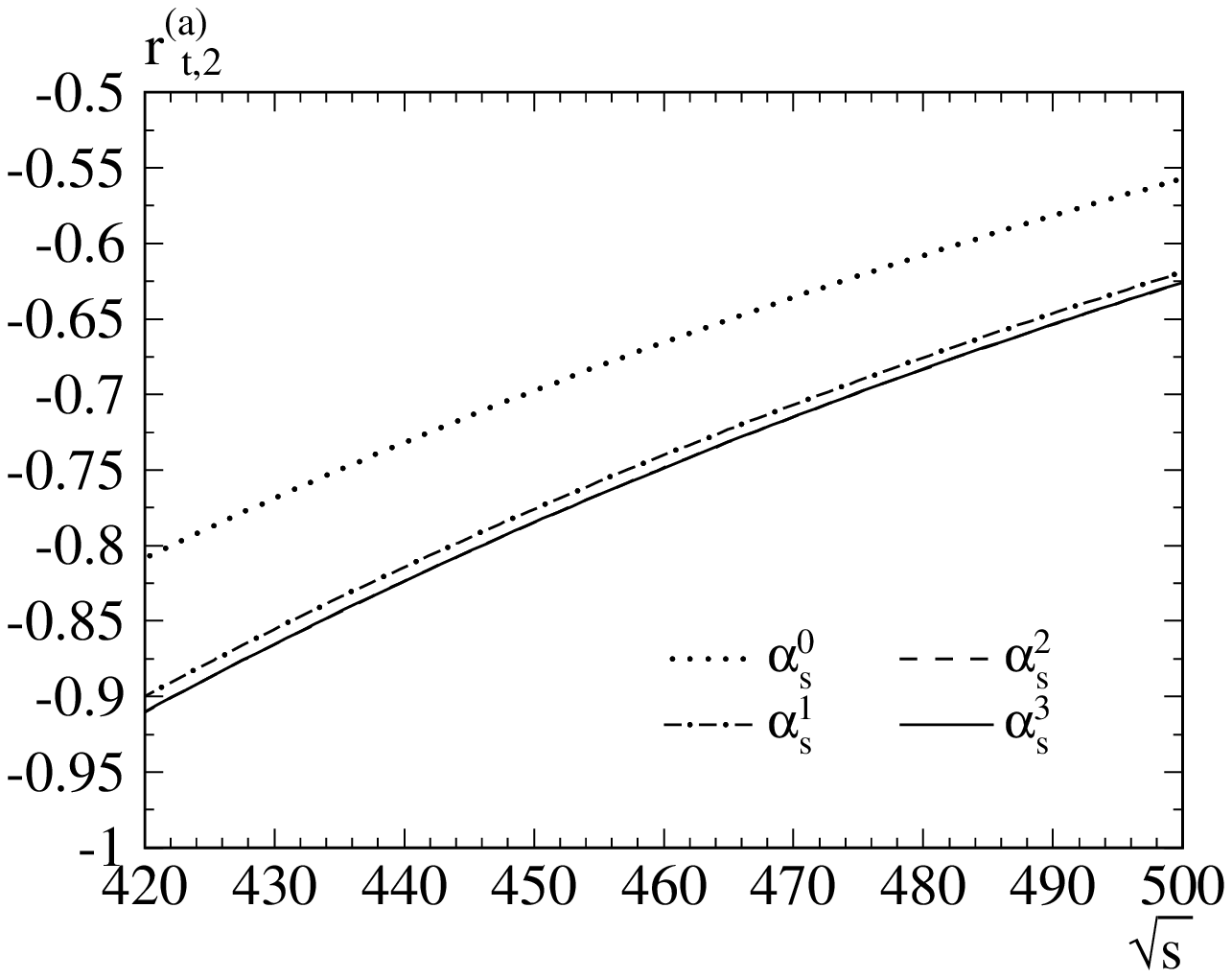}
    \end{tabular}
    \parbox{\captionwidth}{
      \caption[]{\label{fig::rQ2} 
        Quadratic mass corrections ($\propto m^2$) to the non-singlet
        contribution of $r_Q$ for $Q=c,b,t$, arising from diagrams where
        the external current couples directly to the massive quark.
        Upper row: vector current for $c$ and $b$ quarks;
        lower row: vector and axial currents for $t$ quark.
        }}
  \end{center}
\end{figure}
%
%
\begin{figure}[p]
  \begin{center}
    \leavevmode
    \begin{tabular}{rr}
      (a) & (b) \\[-.5em]
      \epsfxsize=18em
      \epsffile[110 265 465 560]{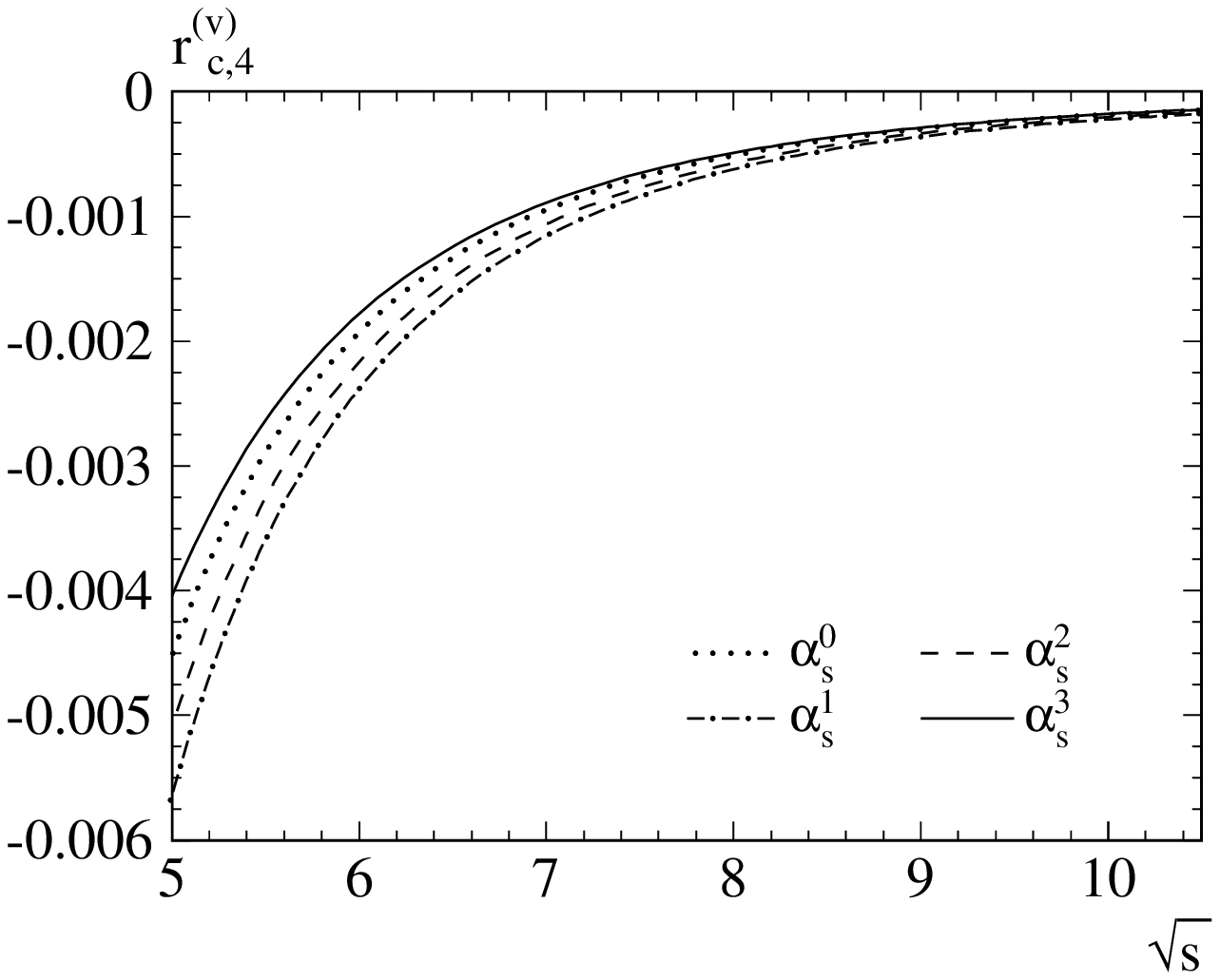} &
      \epsfxsize=18em 
      \epsffile[110 265 465 560]{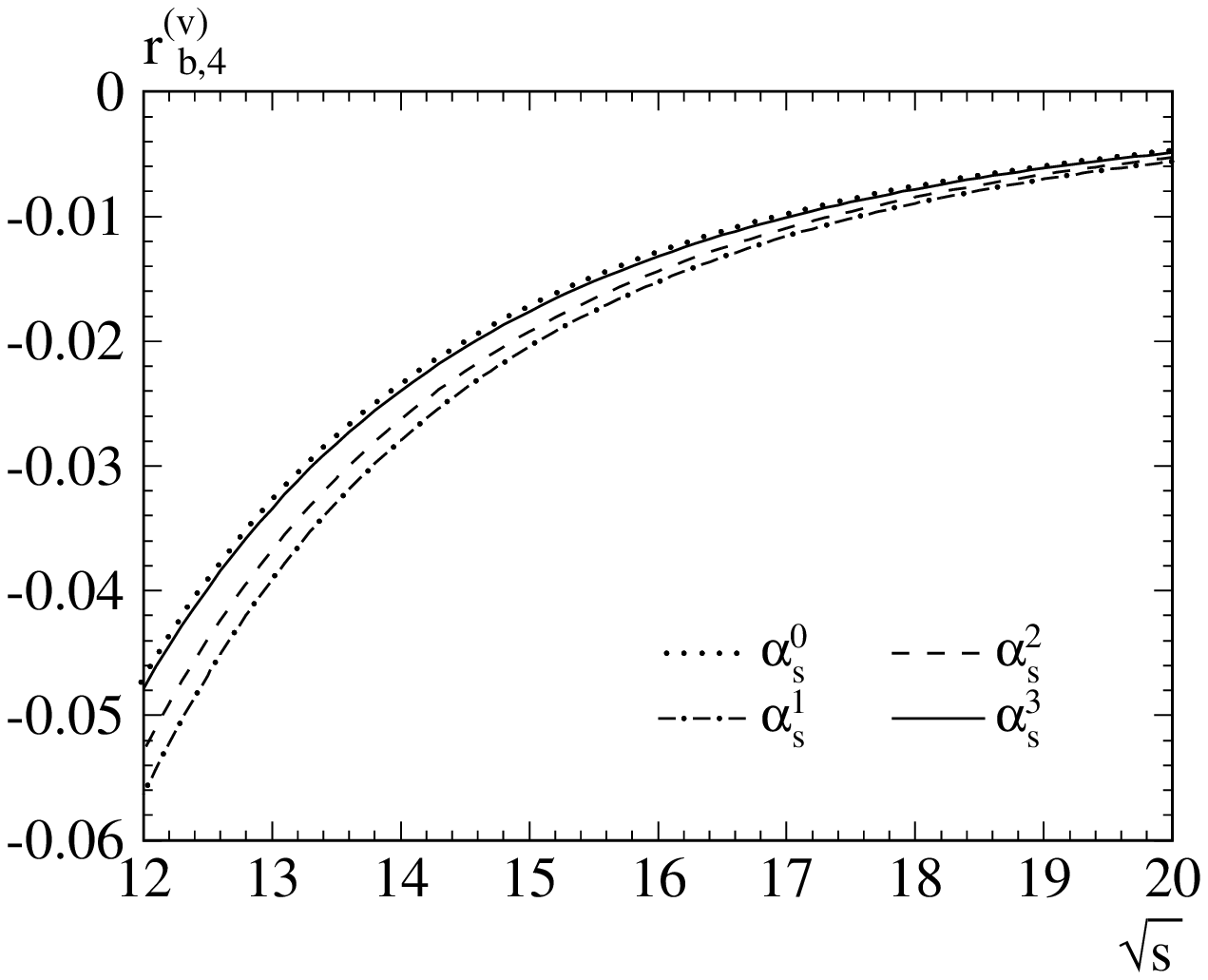} \\[.5em]
      (c) & (d) \\[-.5em]
      \epsfxsize=18em
      \epsffile[110 265 465 560]{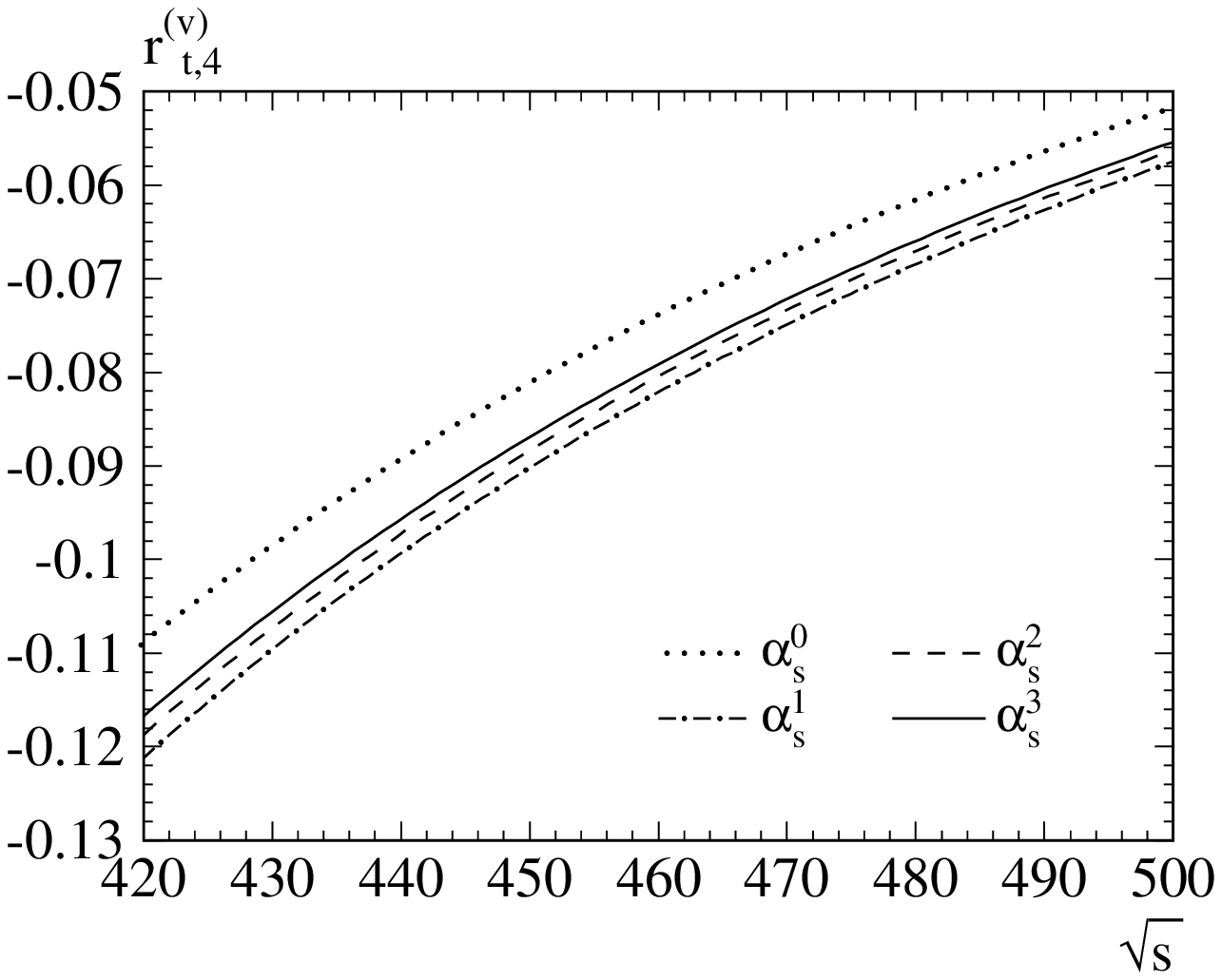} &
      \epsfxsize=18em
      \epsffile[110 265 465 560]{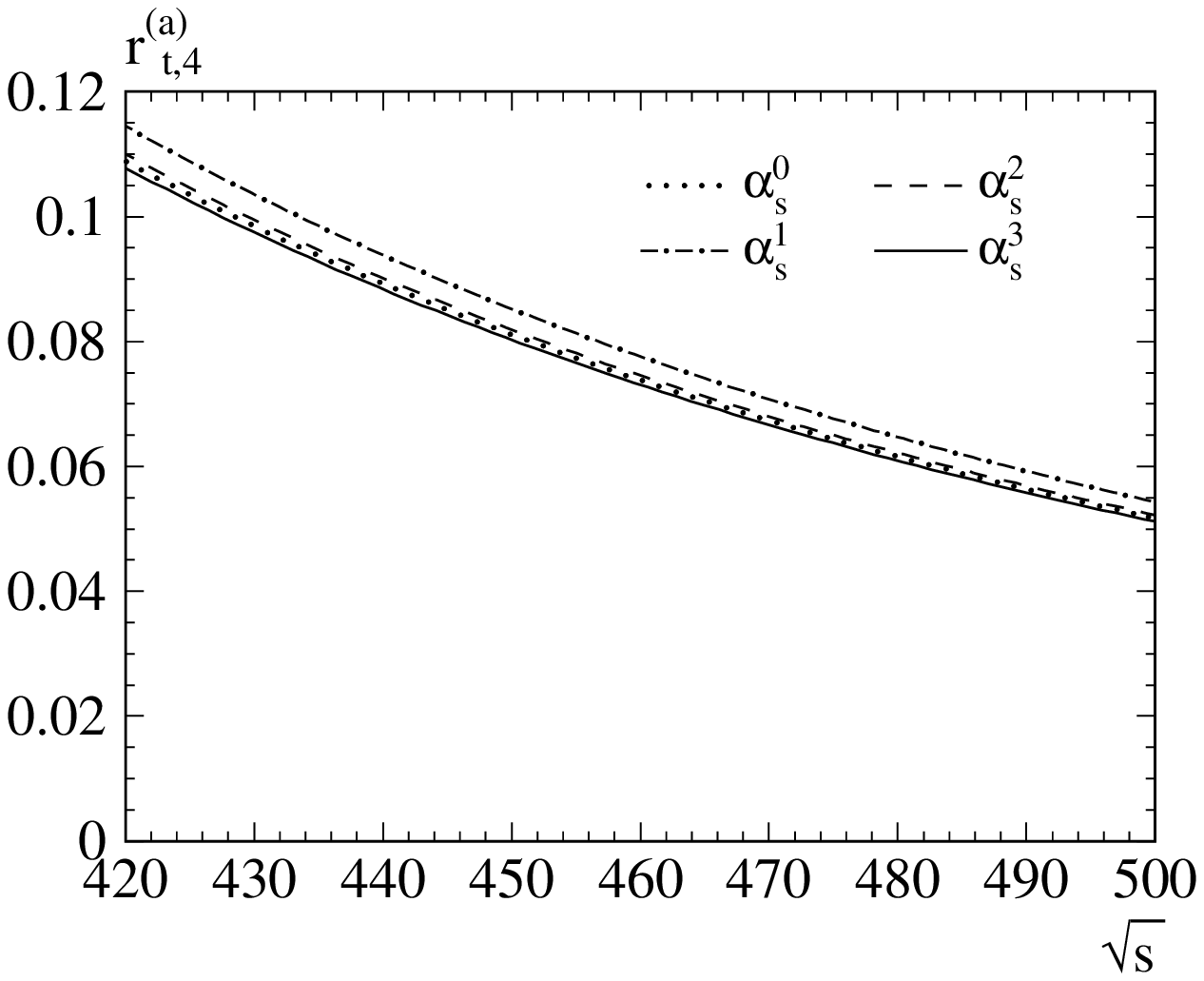}
    \end{tabular}
    \parbox{\captionwidth}{
      \caption[]{\label{fig::rQ4} 
        Quartic mass corrections ($\propto m^4$) to the non-singlet
        contribution of $r_Q$ for $Q=c,b,t$, arising from diagrams where
        the external current couples directly to the massive quark.
        Upper row: vector currents for $c$ and $b$ quarks;
        lower row: vector and axial currents for $t$ quark.
        }}
  \end{center}
\end{figure}
%

%
\begin{figure}[p]
  \begin{center}
    \leavevmode
    \begin{tabular}{rr}
      (a) & (b) \\[-.5em]
      \epsfxsize=18em
      \epsffile[110 265 465 560]{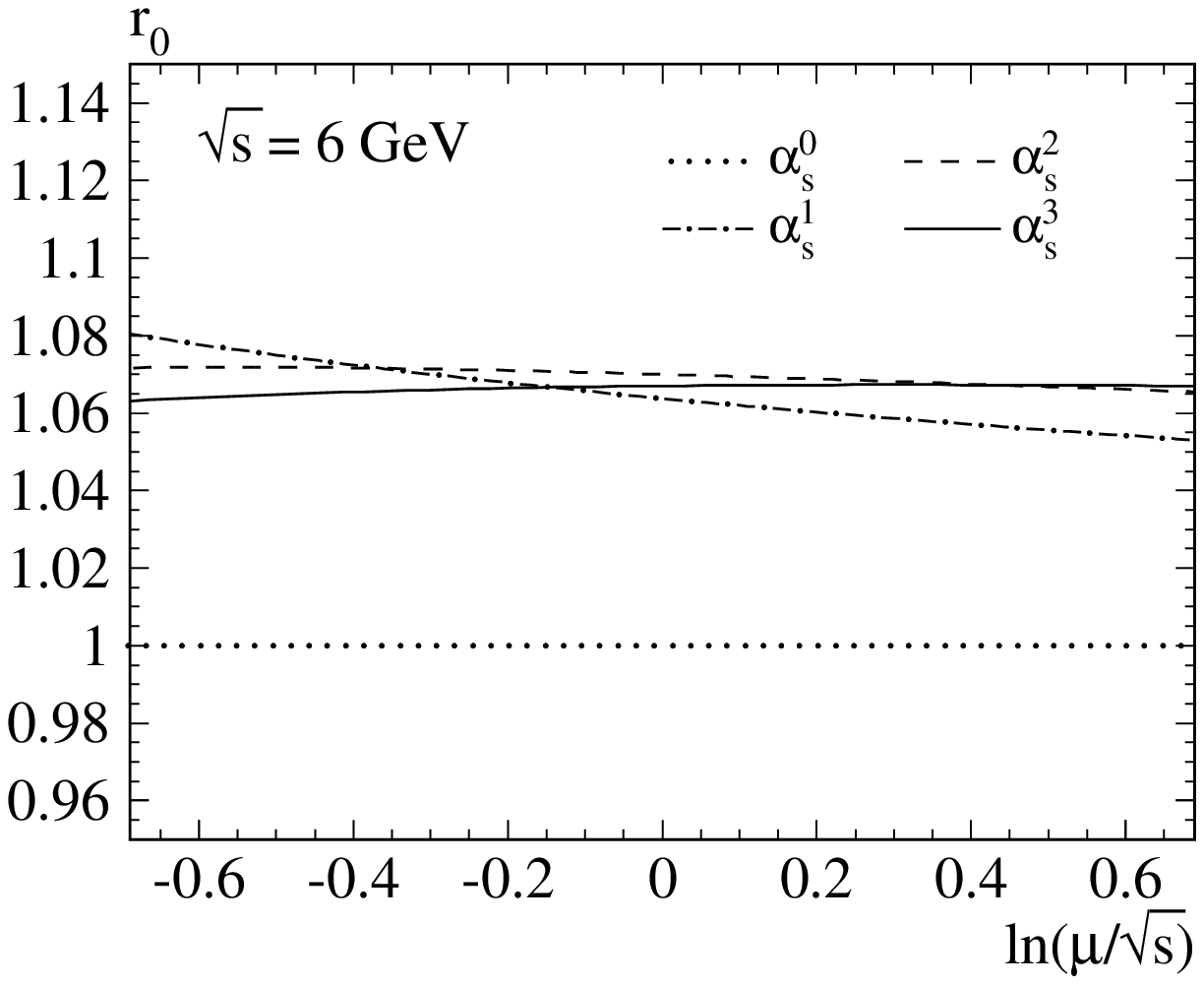} &
      \epsfxsize=18em 
      \epsffile[110 265 465 560]{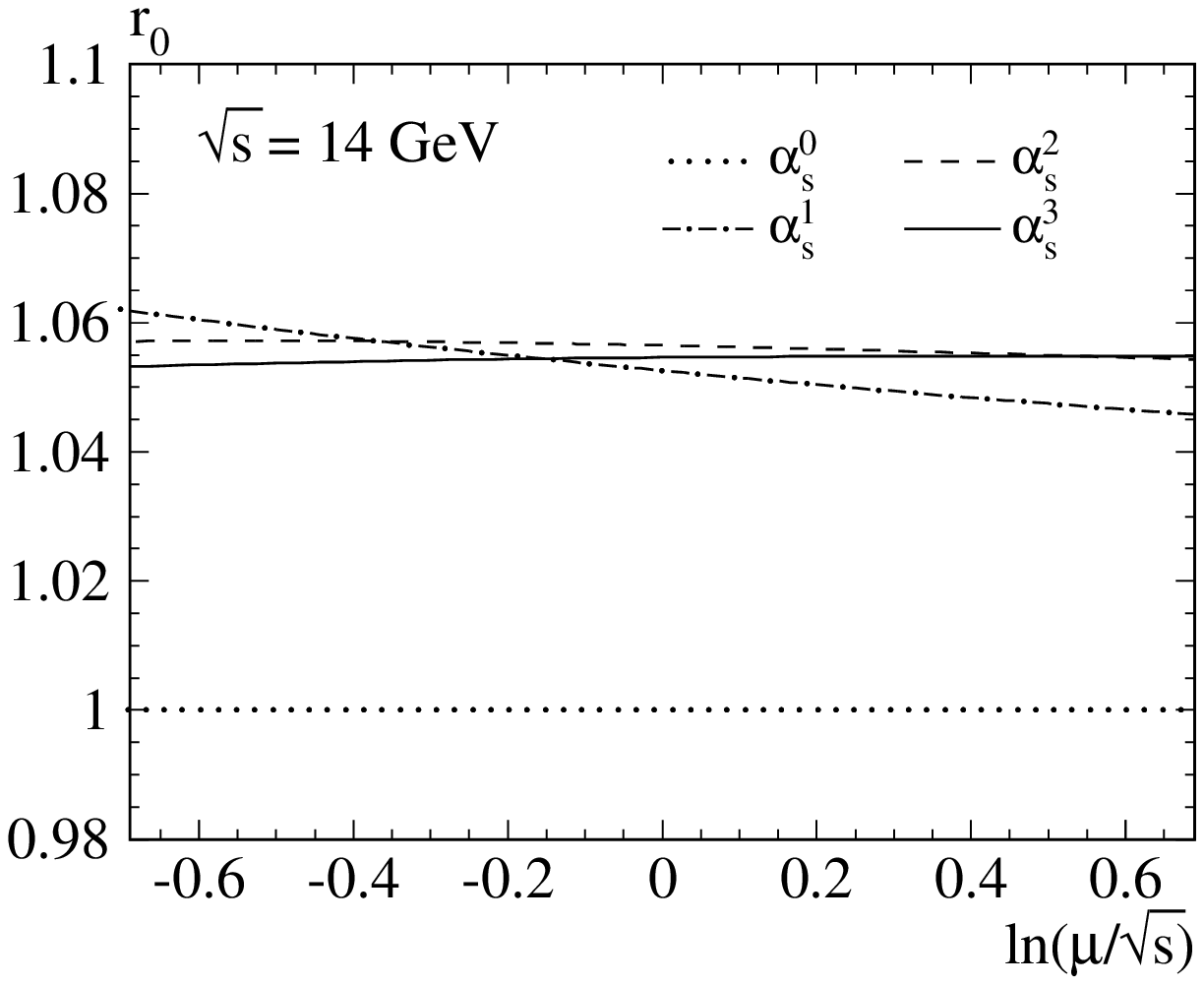} \\[.5em]
       & \multicolumn{1}{c}{\hspace{-2em}(c)}\\[-.5em]
      \multicolumn{2}{c}{
      \epsfxsize=18em
      \epsffile[110 265 465 560]{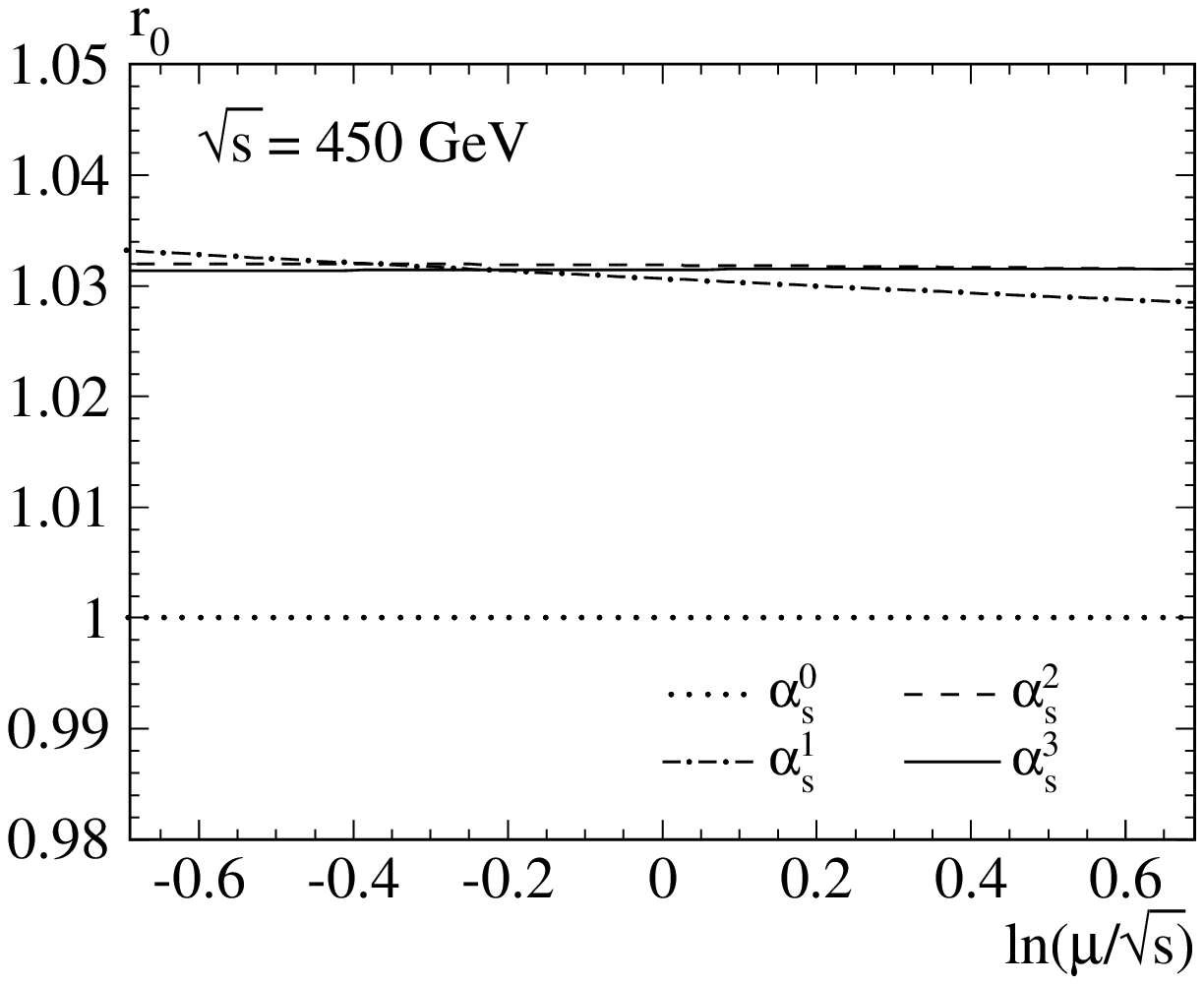}}
    \end{tabular}
    \parbox{\captionwidth}{
      \caption[]{\label{fig::mudep0}
        Variation of $r_0$ with $\mu$. $s$ is fixed to values relevant
        for the production of (a) charm, (b) bottom, and (c) top quarks.
        The abscissa ranges from $\mu=\sqrt{s}/2$ to $\mu=2\sqrt{s}$.
}}
  \end{center}
\end{figure}
%

%
\begin{figure}[p]
  \begin{center}
    \leavevmode
    \begin{tabular}{rr}
      (a) & (b) \\[-.5em]
      \epsfxsize=18em
      \epsffile[110 265 465 560]{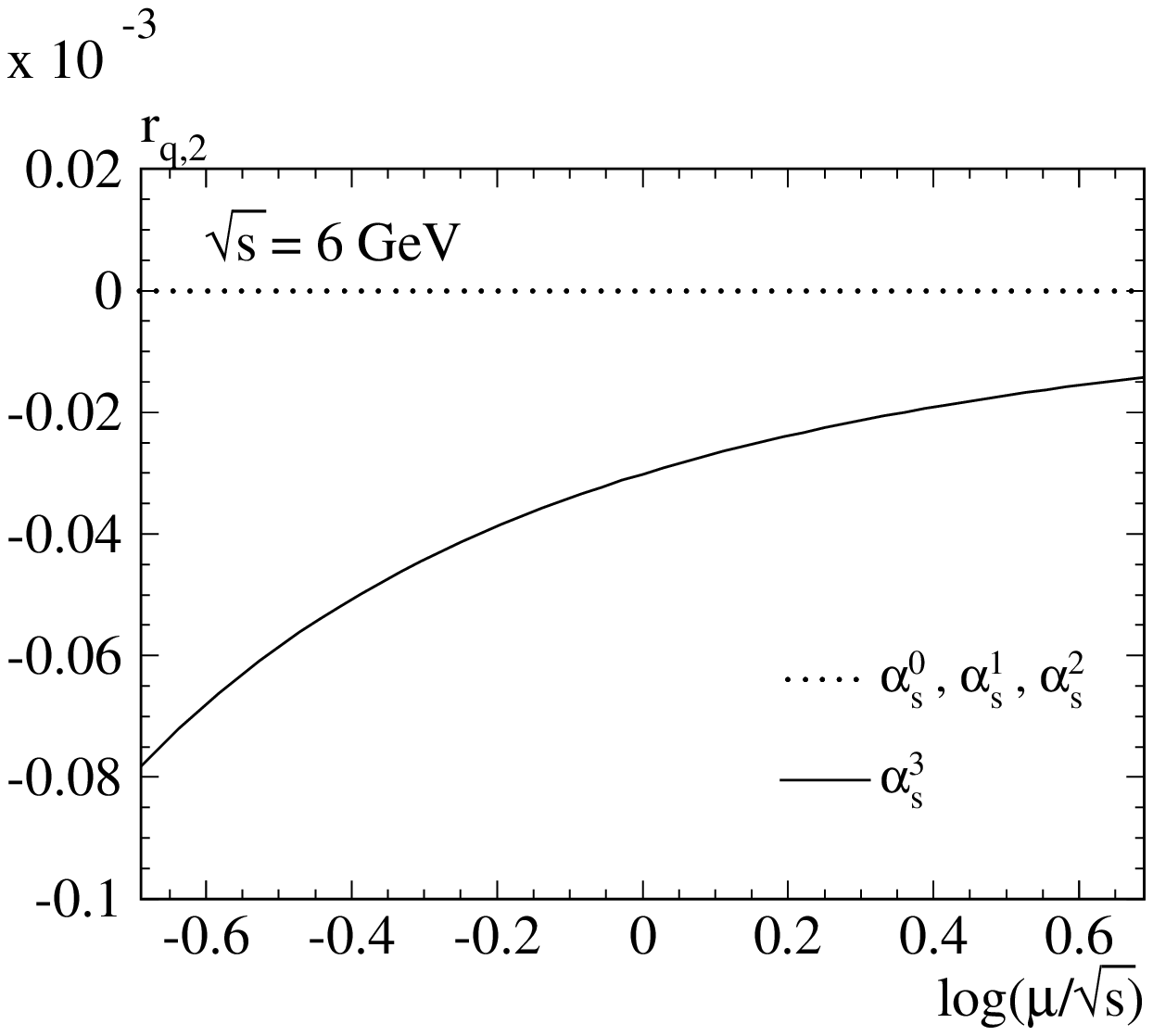} &
      \epsfxsize=18em 
      \epsffile[110 265 465 560]{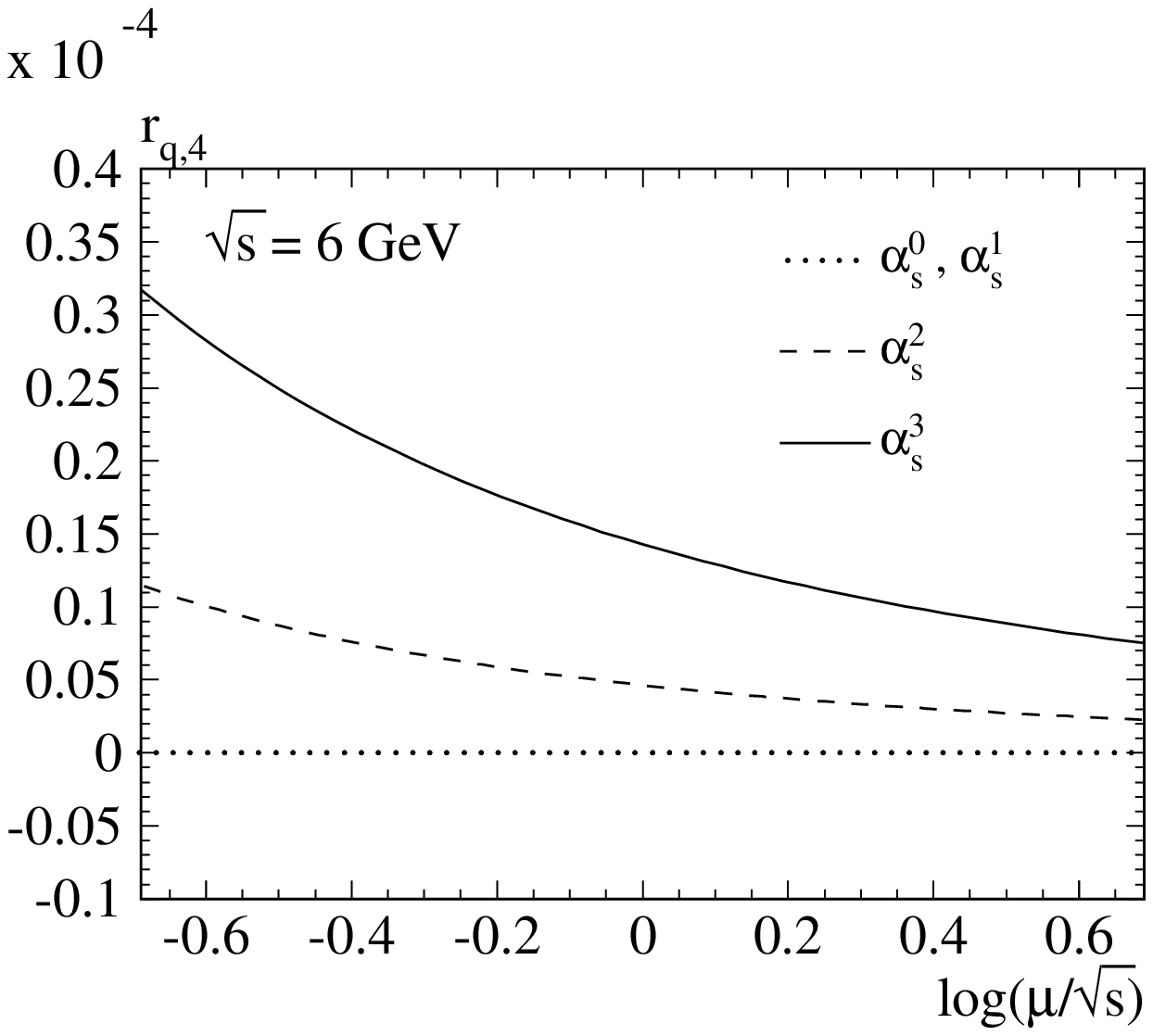} \\[.5em]
      (c) & (d) \\[-.5em]
      \epsfxsize=18em
      \epsffile[110 265 465 560]{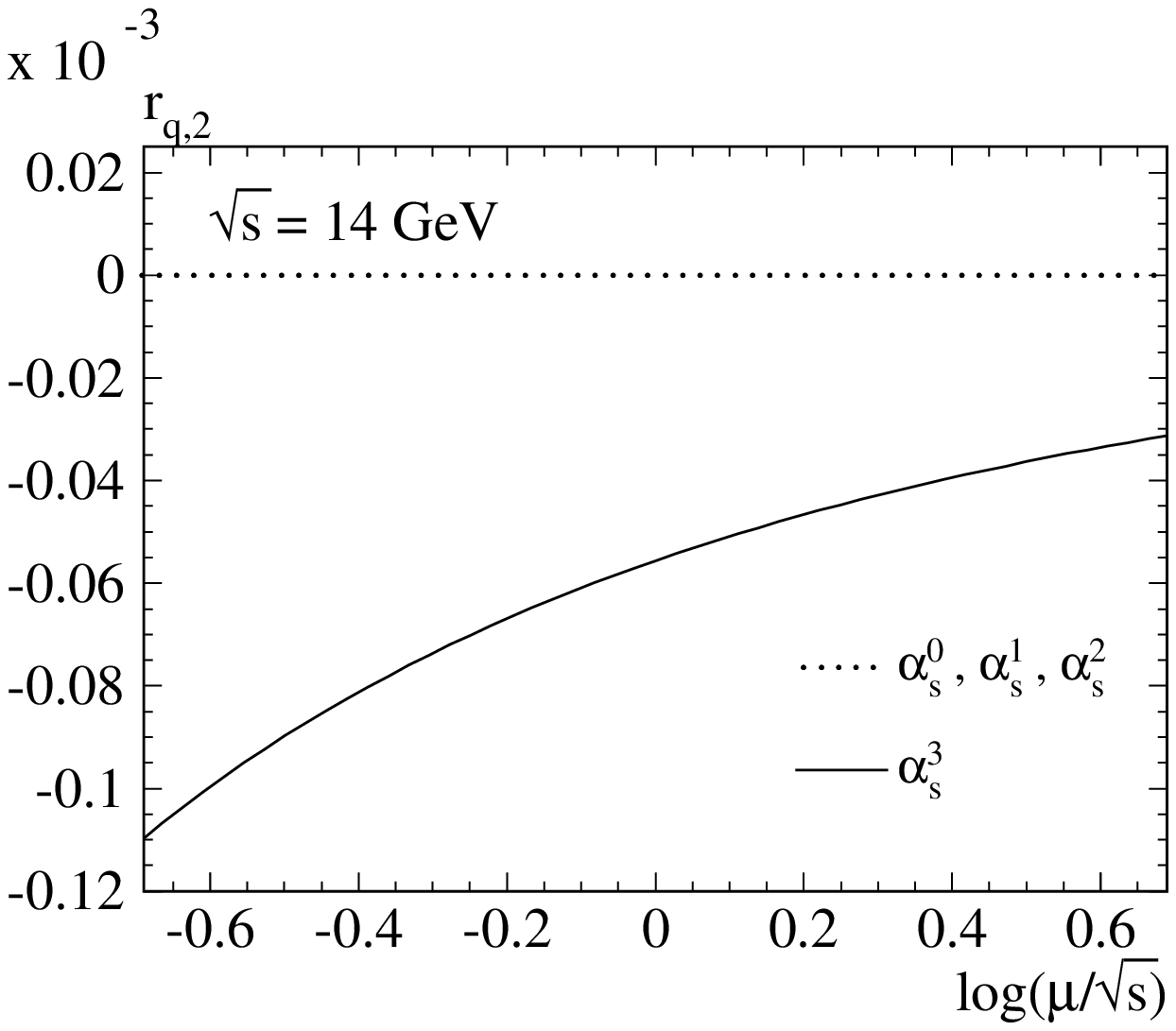} &
      \epsfxsize=18em
      \epsffile[110 265 465 560]{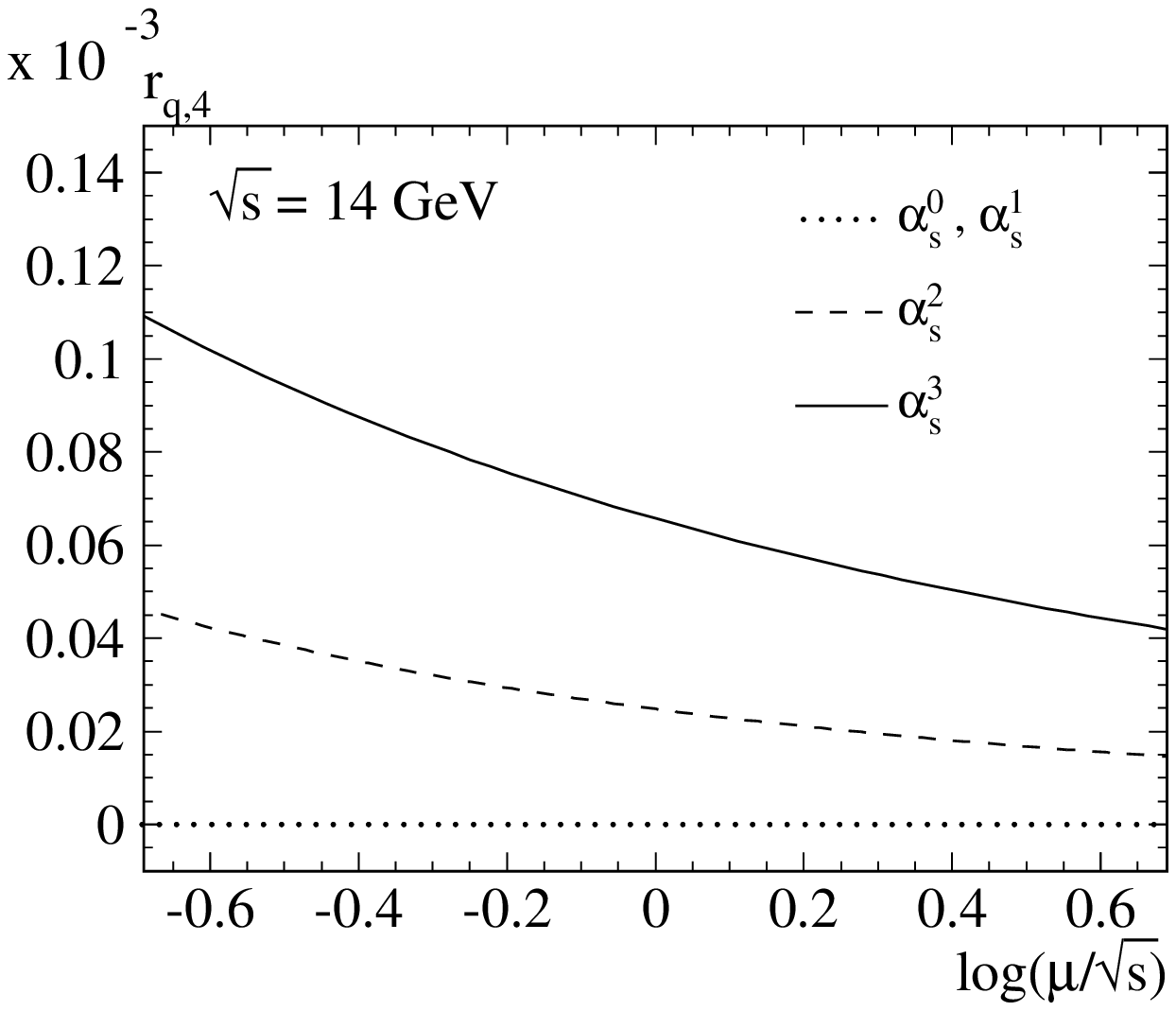} \\[.5em]
      (e) & (f) \\[-.5em]
      \epsfxsize=18em
      \epsffile[110 265 465 560]{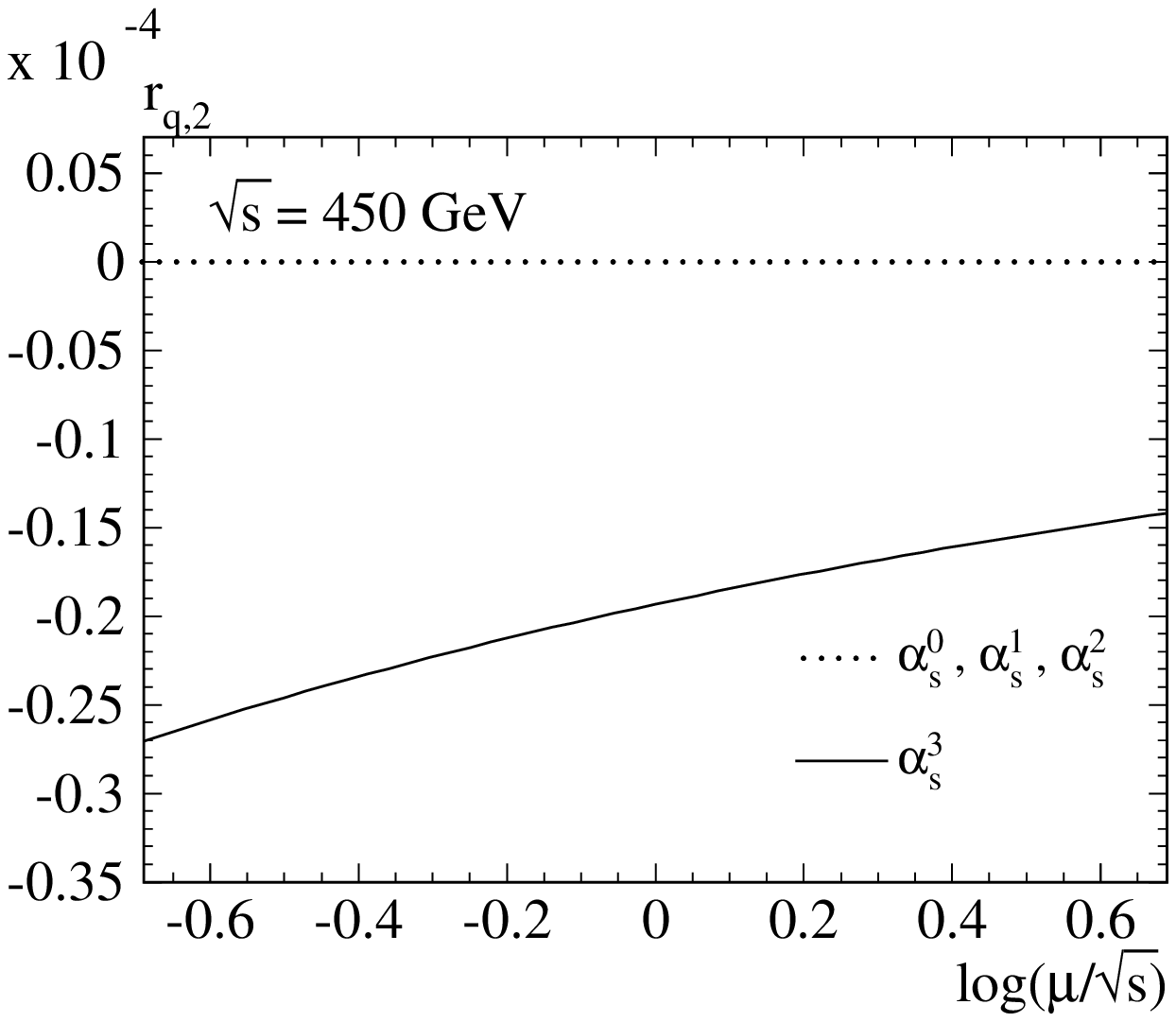} &
      \epsfxsize=18em
      \epsffile[110 265 465 560]{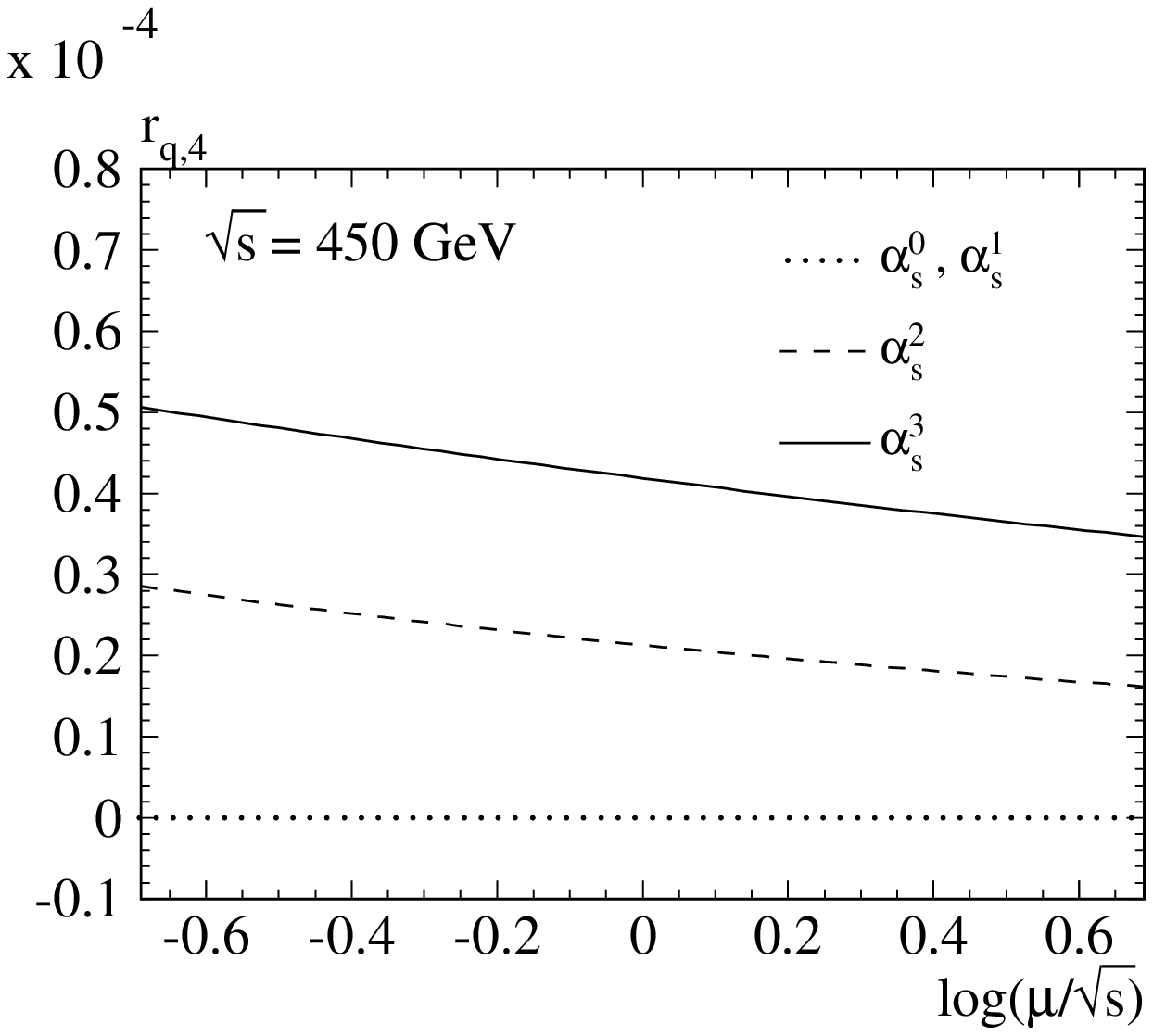}
    \end{tabular}
    \parbox{\captionwidth}{
      \caption[]{\label{fig::mudepv0}
        Variation of $r_{q,2}$ (left column) and $r_{q,4}$ (right
        column) with $\mu$. The first, second, and third row correspond
        to charm, bottom, and top production, respectively.
        }}
  \end{center}
\end{figure}
%
%
\begin{figure}[p]
  \begin{center}
    \leavevmode
    \begin{tabular}{rr}
      (a) & (b) \\[-.5em]
      \epsfxsize=18em
      \epsffile[110 265 465 560]{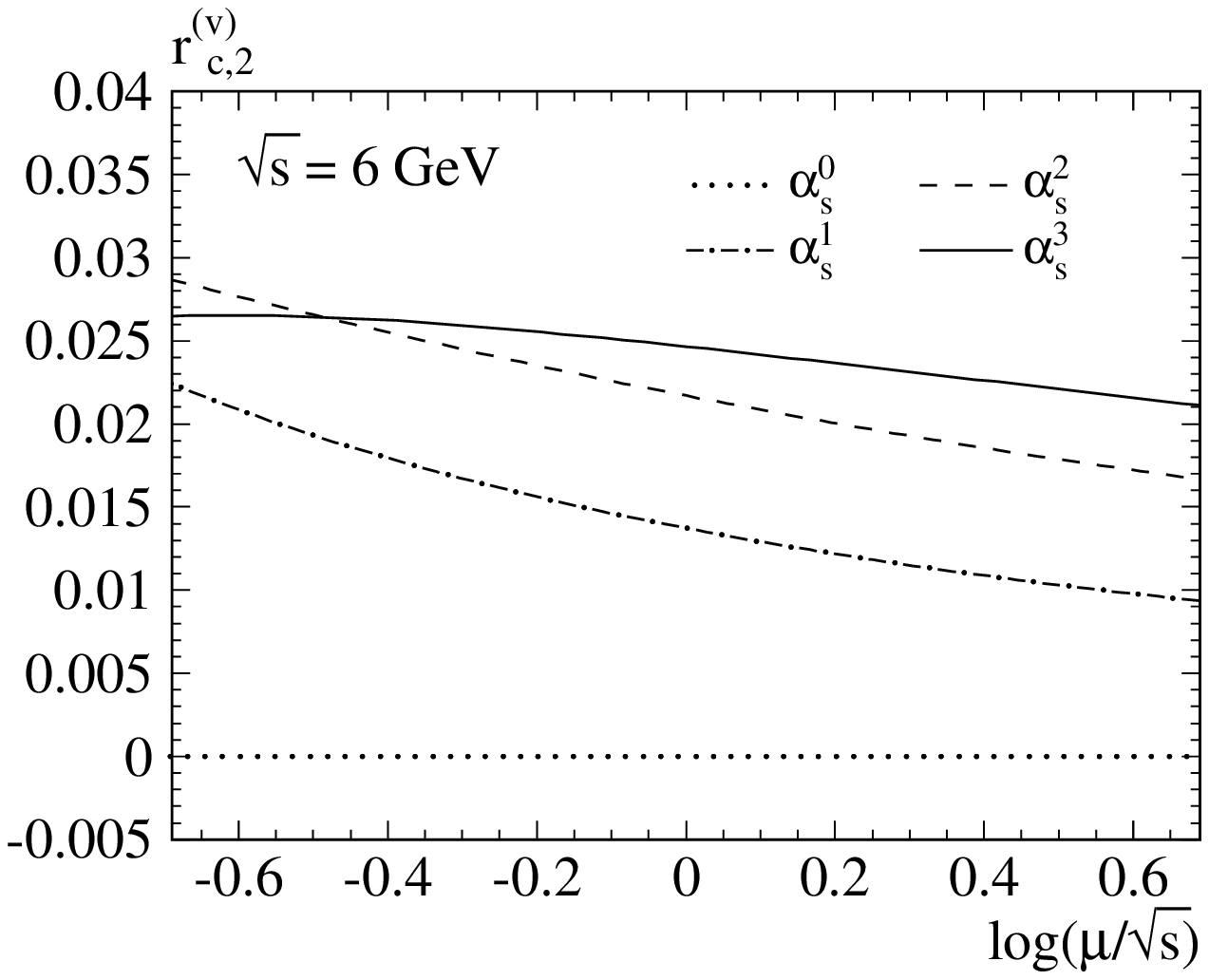} &
      \epsfxsize=18em 
      \epsffile[110 265 465 560]{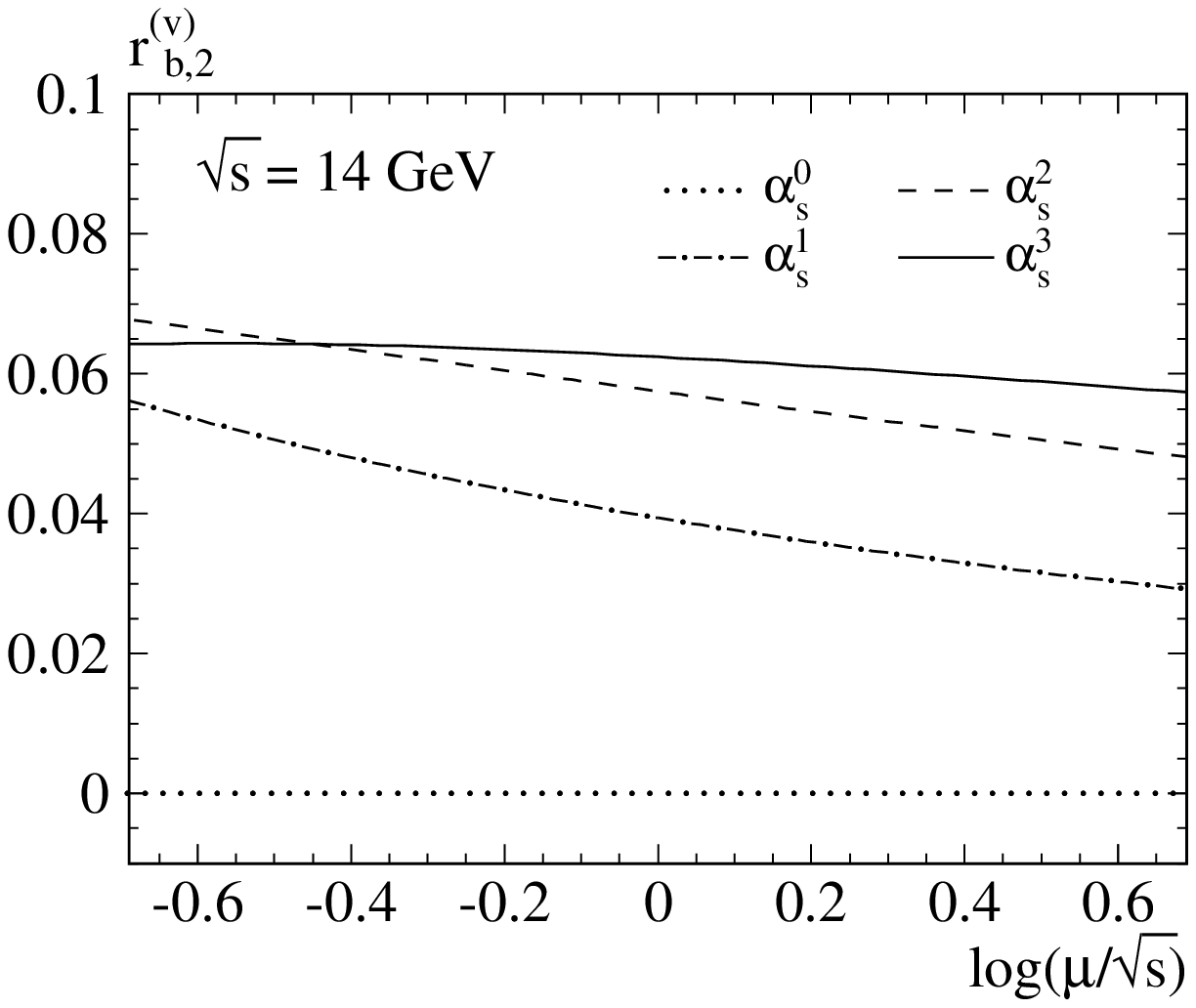} \\[.5em]
      (c) & (d) \\[-.5em]
      \epsfxsize=18em
      \epsffile[110 265 465 560]{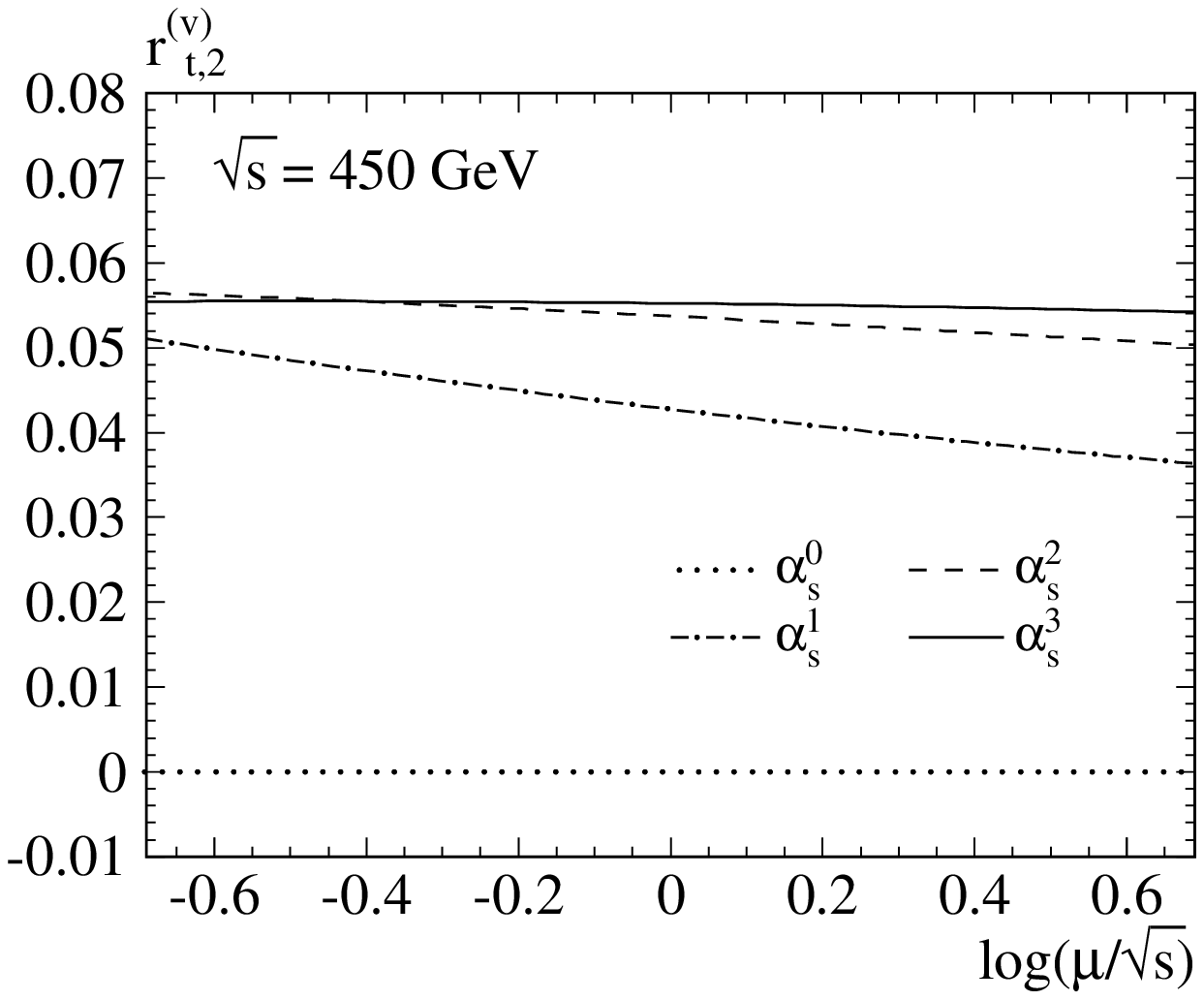} &
      \epsfxsize=18em
      \epsffile[110 265 465 560]{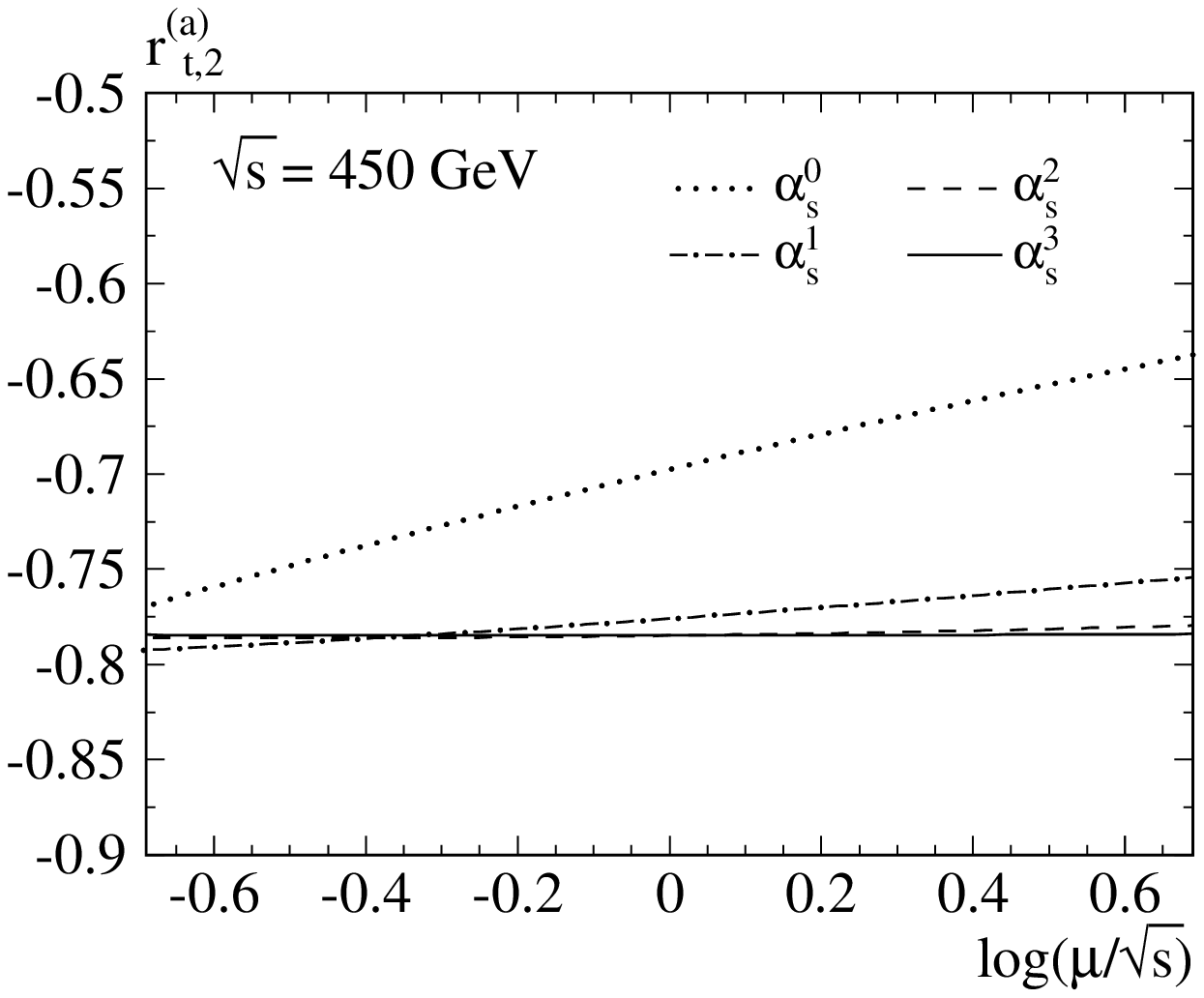}
    \end{tabular}
    \parbox{\captionwidth}{
      \caption[]{\label{fig::mudep2}
        Variation of $r^{(v/a)}_{Q,2}$ with $\mu$.
        First row: (a) charm, (b) bottom production for vector case;
        second row: (c) vector and (d) axial-vector case for top production.
        }}
  \end{center}
\end{figure}
%
%
\begin{figure}[p]
  \begin{center}
    \leavevmode
    \begin{tabular}{rr}
      (a) & (b) \\[-.5em]
      \epsfxsize=18em
      \epsffile[110 265 465 560]{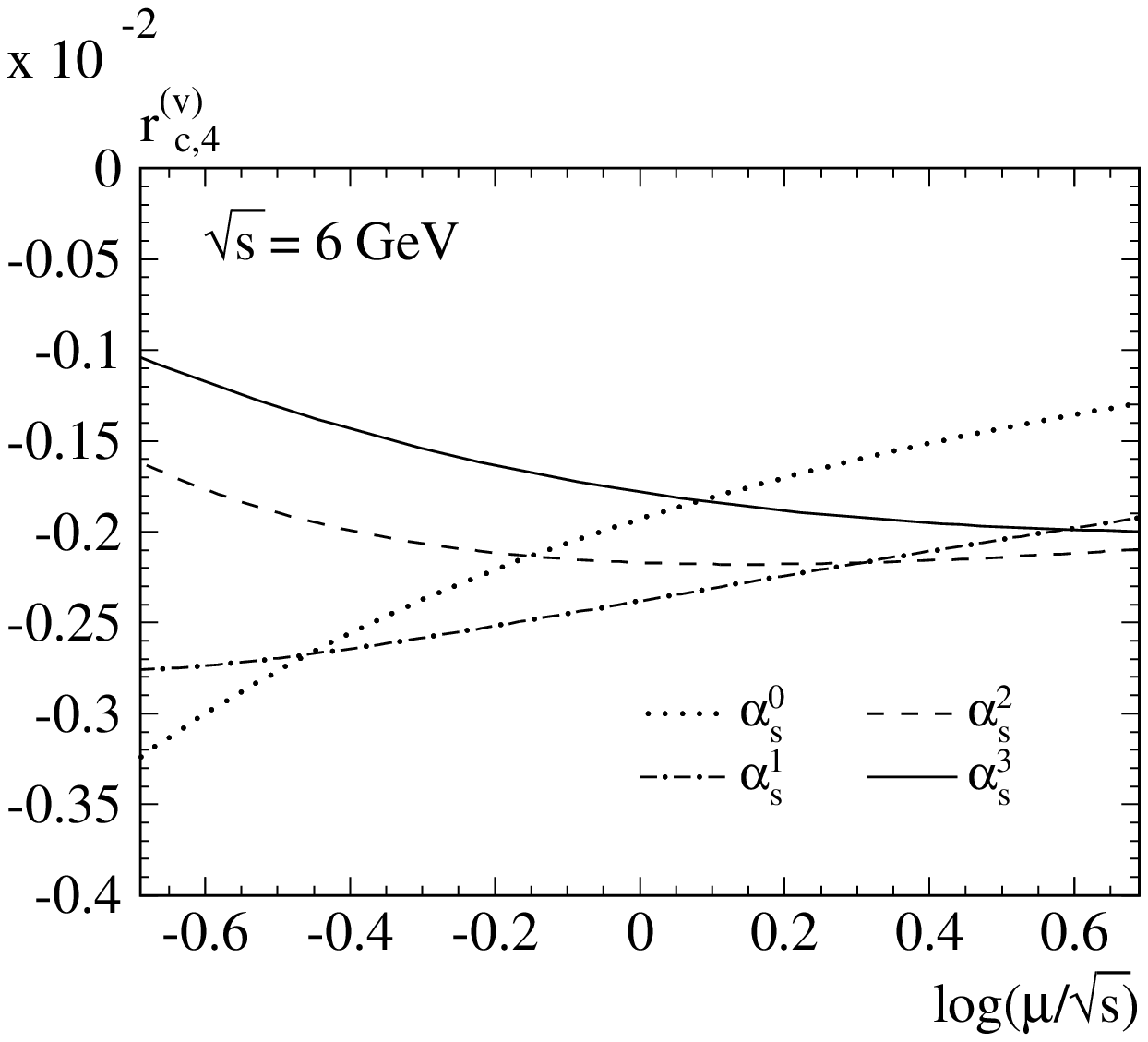} &
      \epsfxsize=18em 
      \epsffile[110 265 465 560]{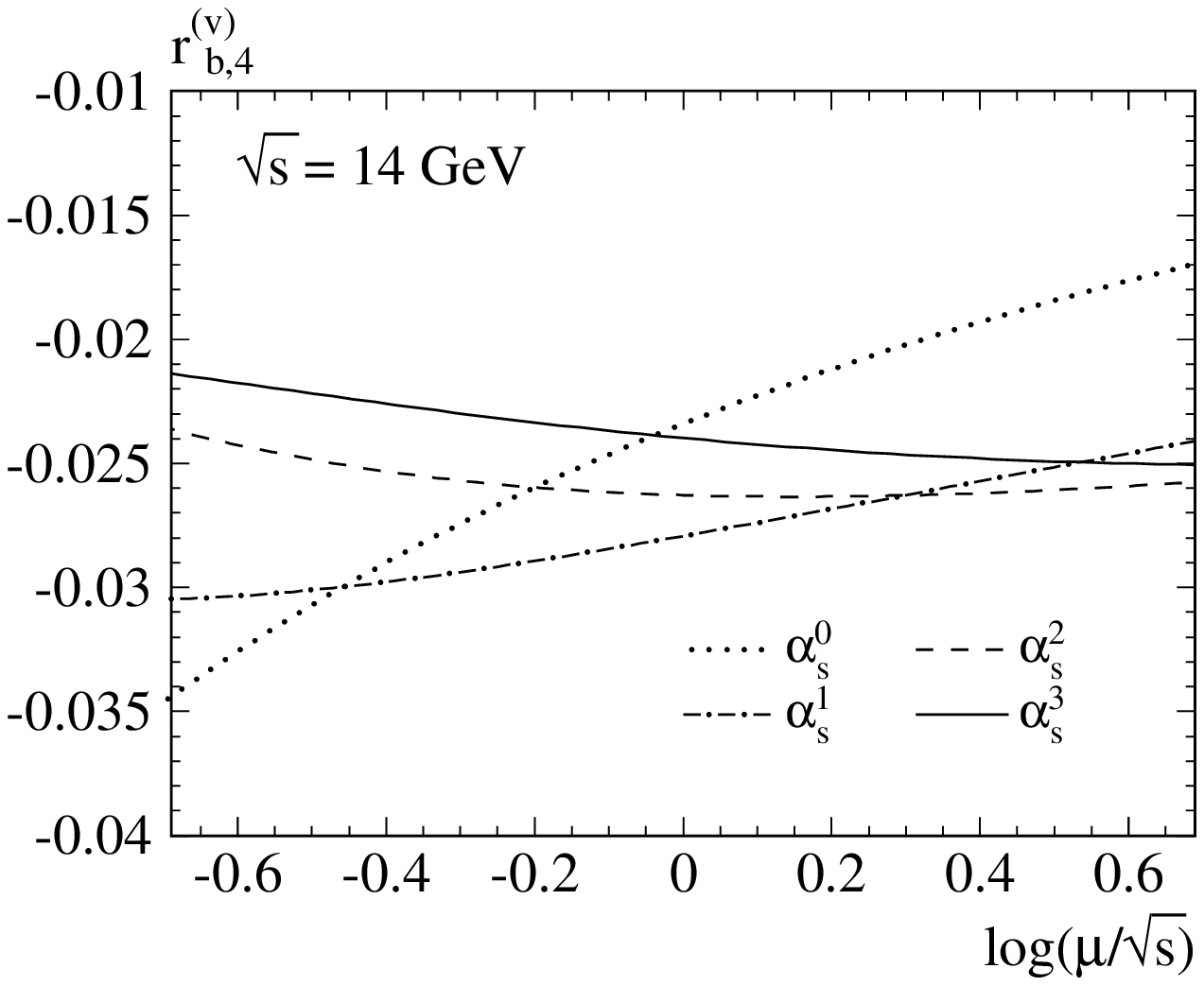} \\[.5em]
      (c) & (d) \\[-.5em]
      \epsfxsize=18em
      \epsffile[110 265 465 560]{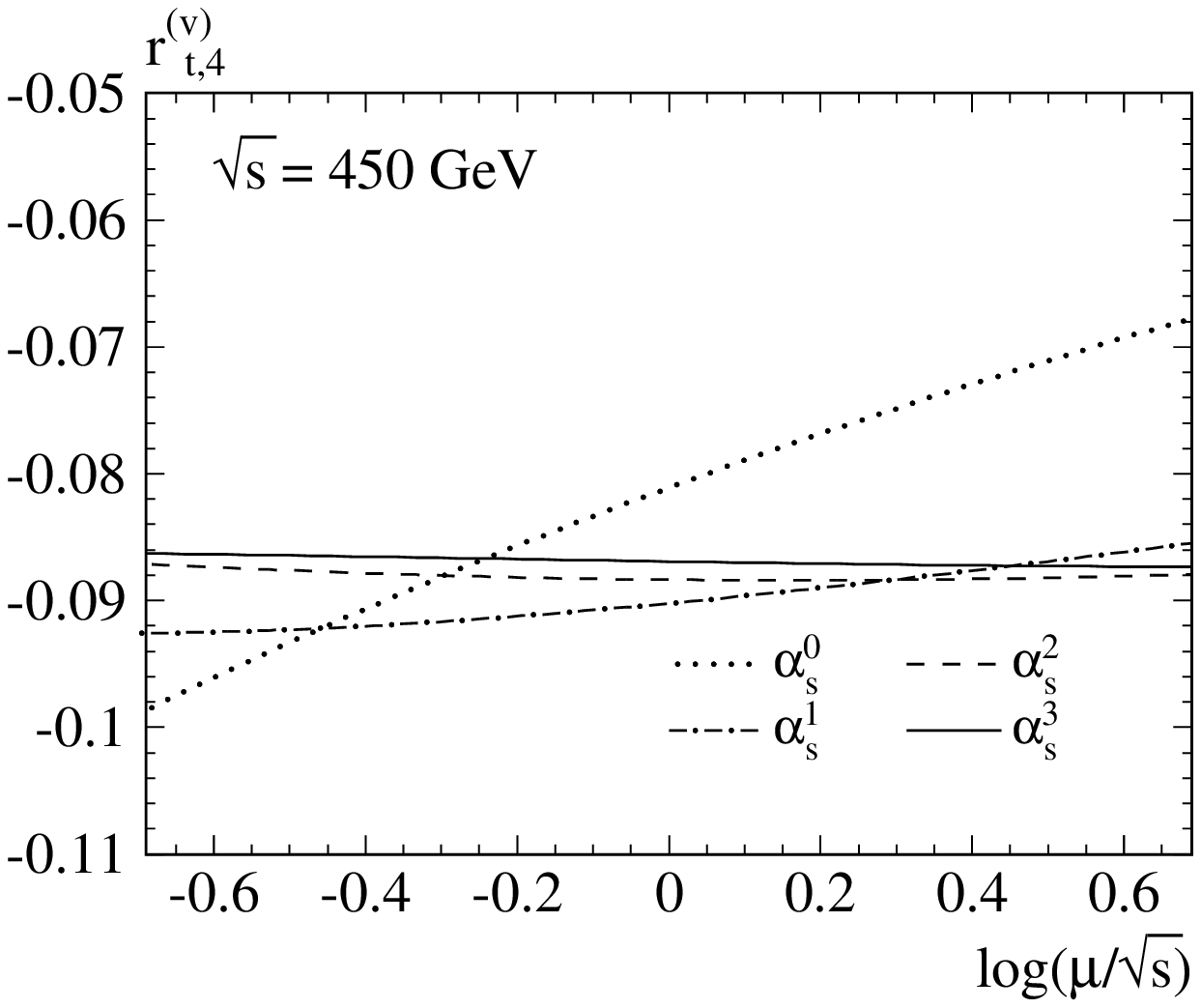} &
      \epsfxsize=18em
      \epsffile[110 265 465 560]{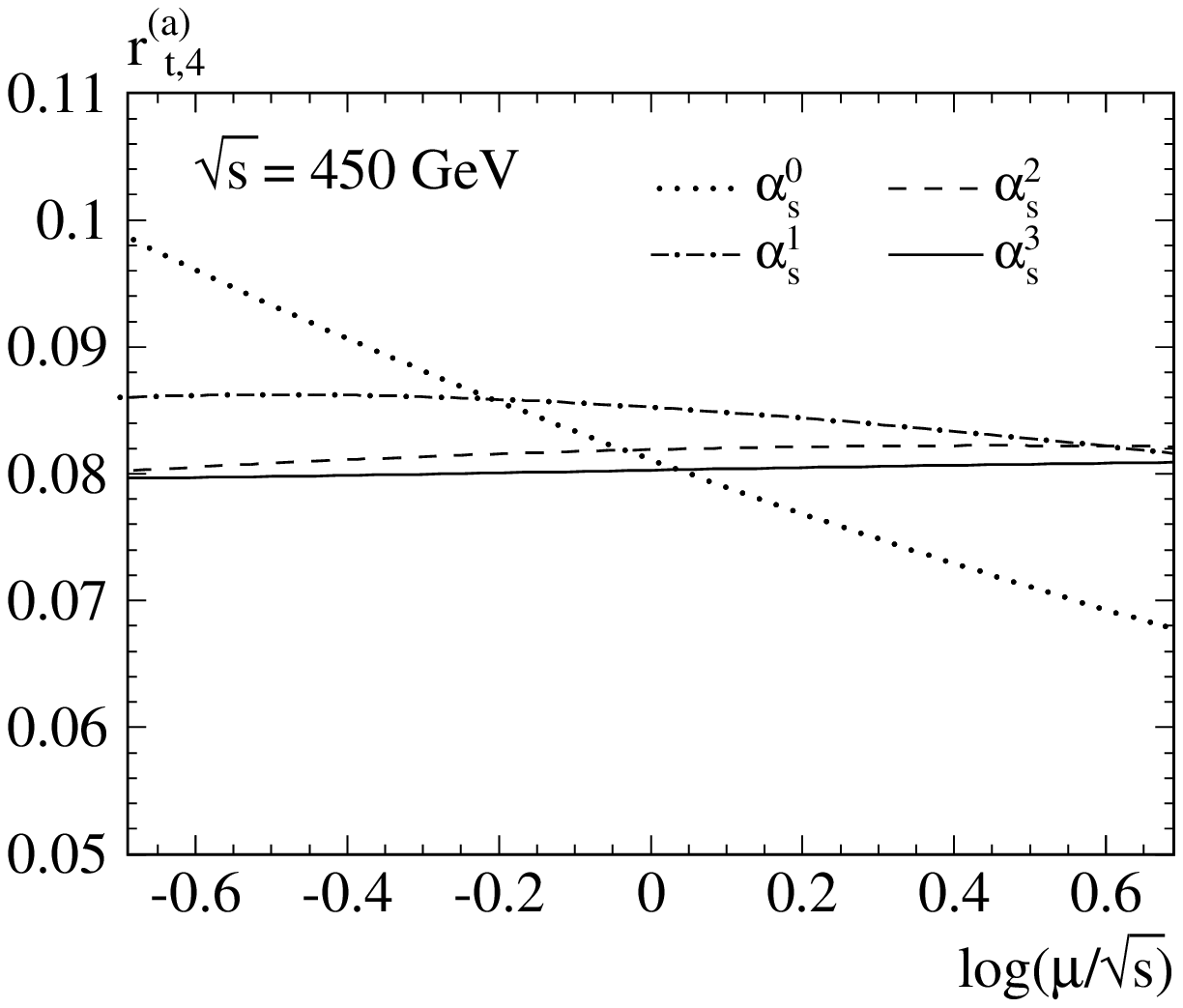}
    \end{tabular}
    \parbox{\captionwidth}{
      \caption[]{\label{fig::mudep4}
        Variation of $r^{(v/a)}_{Q,4}$ with $\mu$.
        Conventions as in Fig.~\ref{fig::mudep2}.
        }}
  \end{center}
\end{figure}
%
%
\begin{figure}[p]
  \begin{center}
    \leavevmode
    \begin{tabular}{rr}
      (a) & (b) \\[-.5em]
      \epsfxsize=18em
      \epsffile[110 265 465 560]{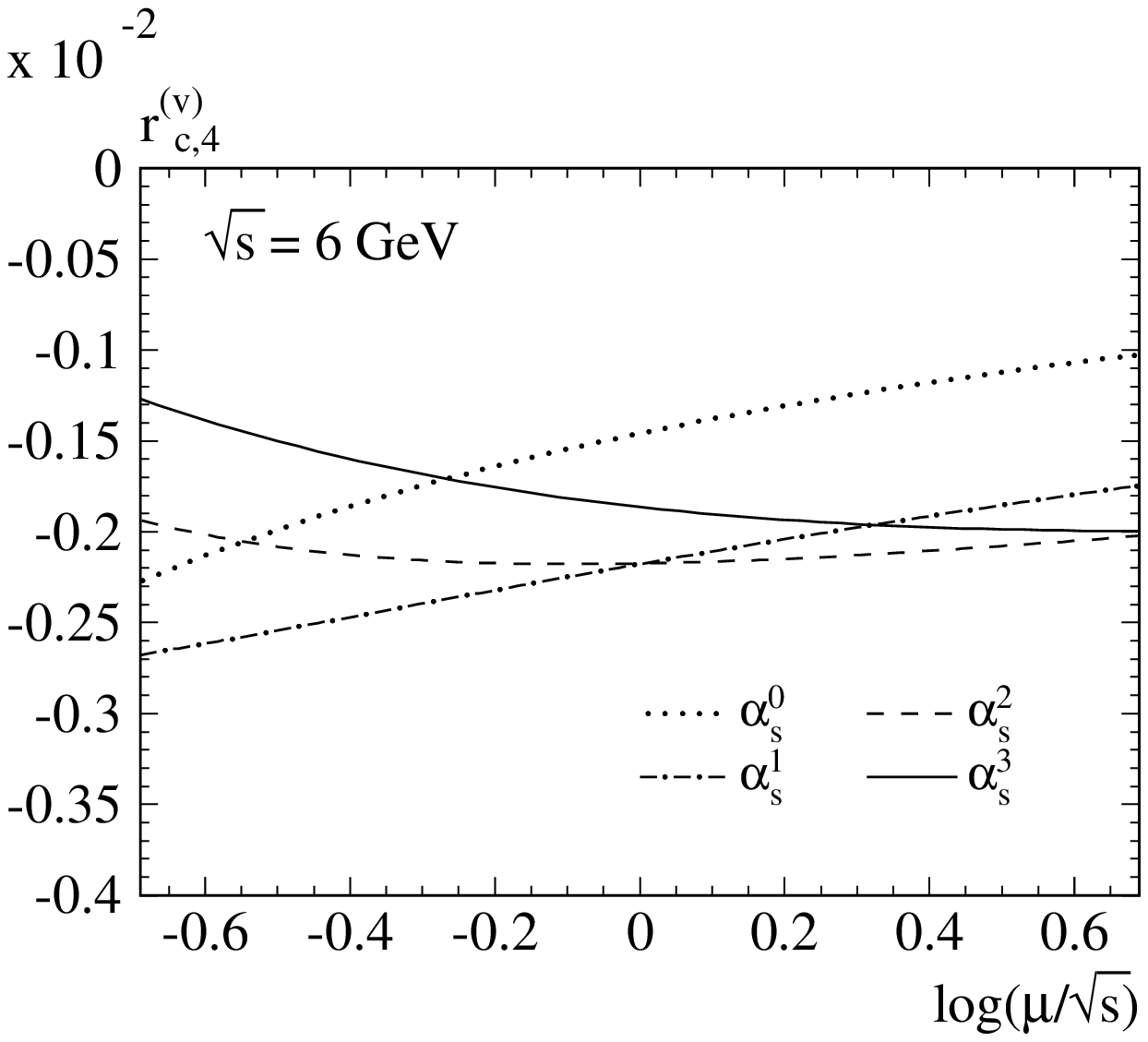} &
      \epsfxsize=18em 
      \epsffile[110 265 465 560]{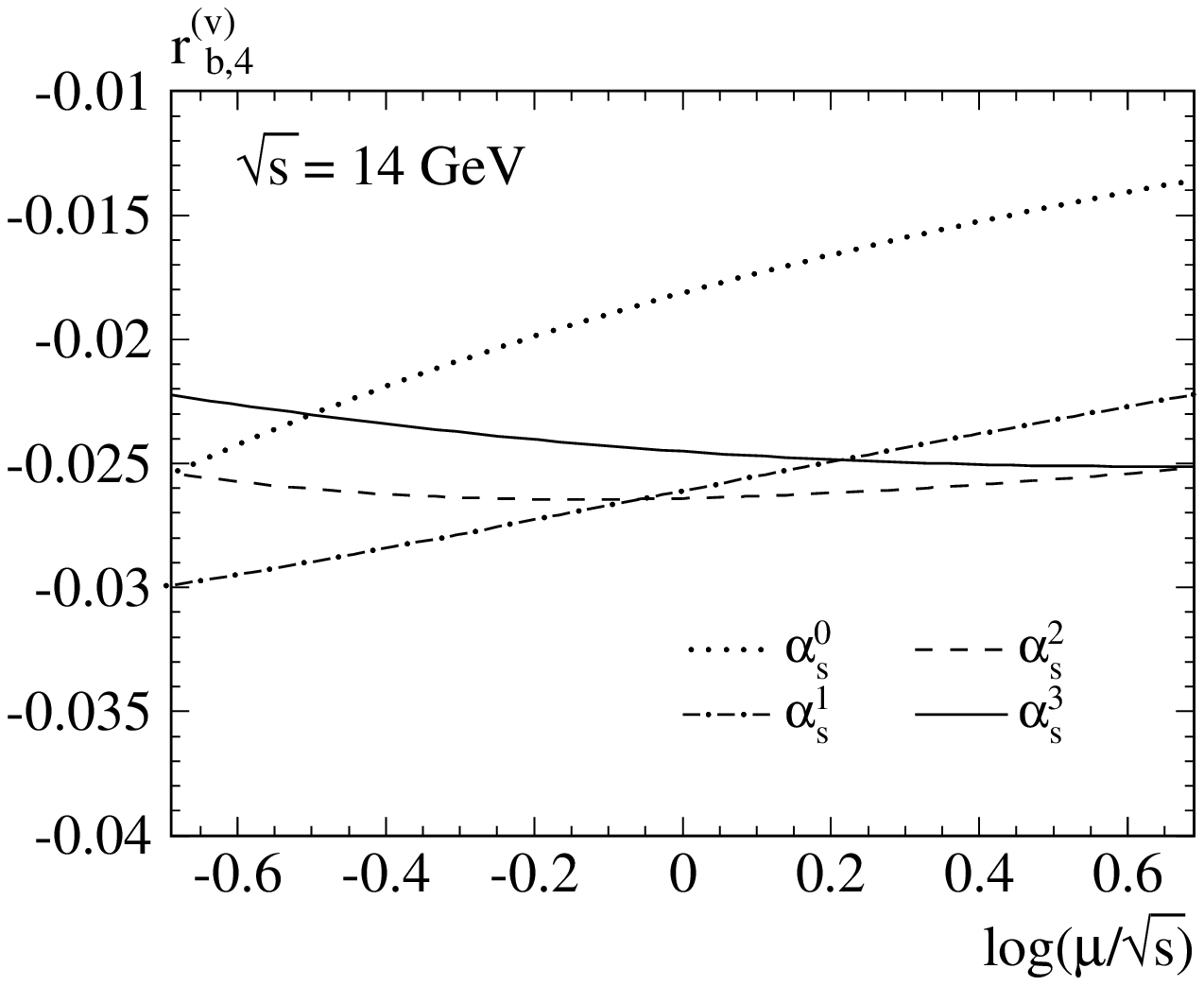} \\[.5em]
      (c) & (d) \\[-.5em]
      \epsfxsize=18em
      \epsffile[110 265 465 560]{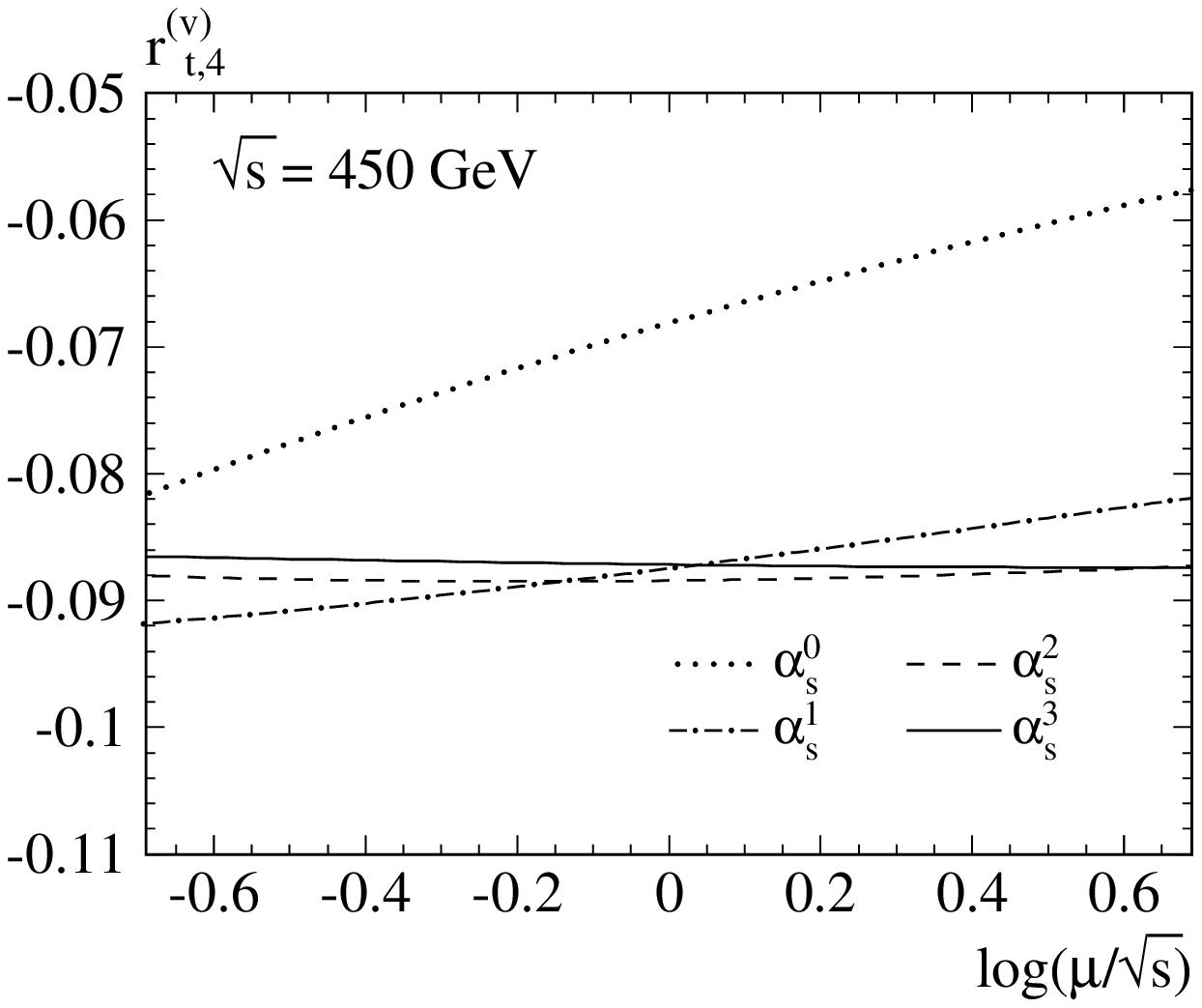} &
      \epsfxsize=18em
      \epsffile[110 265 465 560]{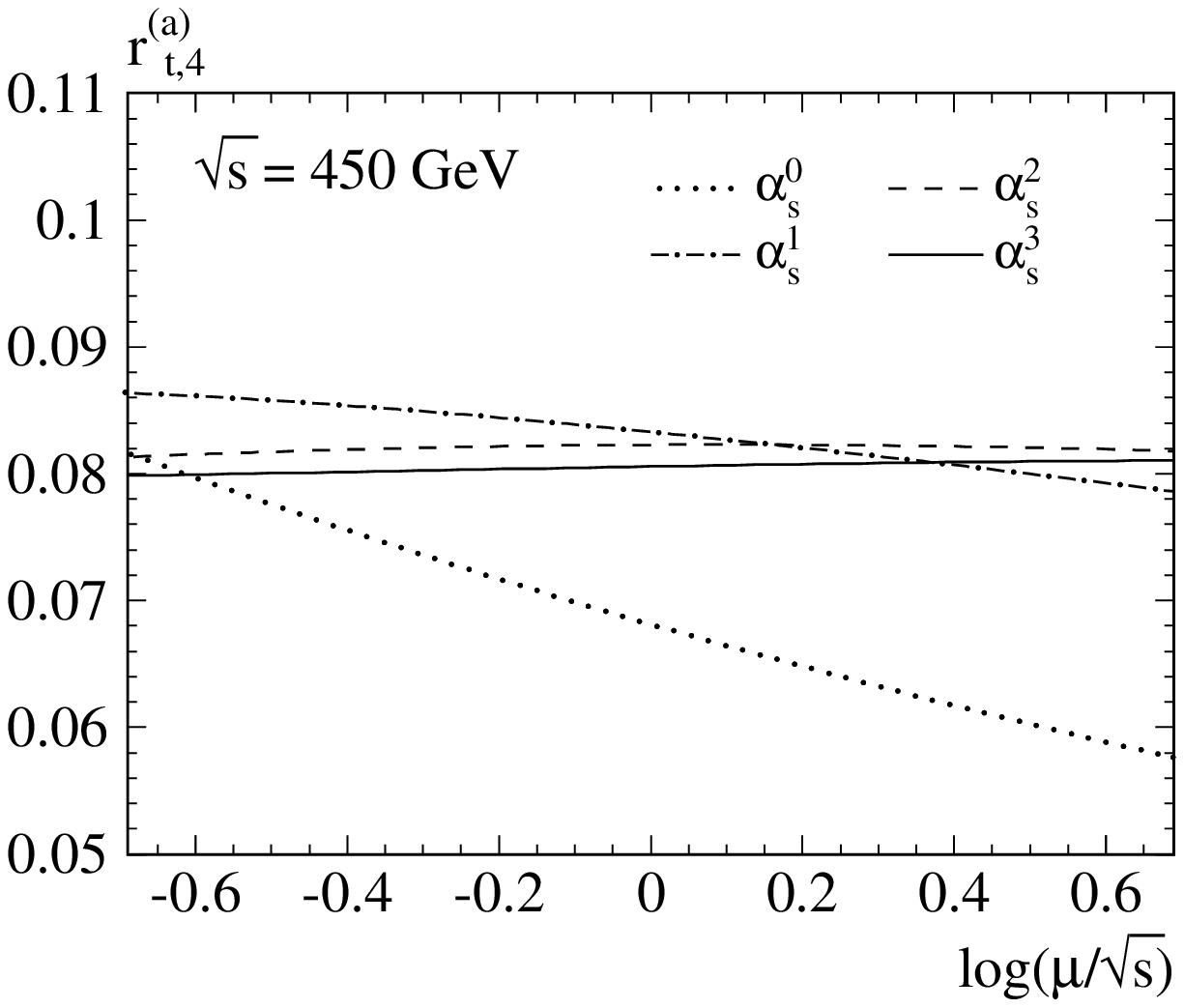}
    \end{tabular}
    \parbox{\captionwidth}{
      \caption[]{\label{fig::mudep4si}
        Variation of $r^{(v/a)}_{Q,4}$ with $\mu$, using the invariant
        mass $\hat m$. The conventions are the same as in
        Fig.~\ref{fig::mudep2} and \ref{fig::mudep4}.
        }}
  \end{center}
\end{figure}
%

\end{document}